\documentstyle[epsfig,a4p,12pt,rotating]{article}
\begin{document}

%
%
\newcommand{\PRnum}    {CERN-EP/99-122}
\newcommand{\PNnum}     {OPAL Physics Note PN-406}
\newcommand{\TNnum}     {OPAL Technical Note TN-xxx}
\newcommand{\Date}      {\today}
\newcommand{\Author}    {D.~I.~Futyan, G.W.~Wilson, T.R.~Wyatt}
\newcommand{\MailAddr}  {david.futyan@cern.ch}
\newcommand{\EdBoard}   {E.~Duchovni S.~Komamiya, R.~,van Kooten, M.~Mannelli.}
\newcommand{\DraftVer}  {Version 1.0}
\newcommand{\DraftDate} {\today}
\newcommand{\TimeLimit} {Comments to david.futyan@cern.ch by Thursday, September 2nd 1999, 12h00 CERN time, please.}

\def\toprule{\noalign{\hrule \medskip}}
\def\midrule{\noalign{\medskip\hrule }}
\def\botrule{\noalign{\medskip\hrule }}
\setlength{\parskip}{\medskipamount}


\newcommand{\ee}{{\mathrm e}^+ {\mathrm e}^-}
\newcommand{\sq}{\tilde{\mathrm q}}
\newcommand{\seff}{\tilde{\mathrm f}}
\newcommand{\sele}{\tilde{\mathrm e}}
\newcommand{\sell}{\tilde{\ell}}
\newcommand{\snu}{\tilde{\nu}}
\newcommand{\smu}{\tilde{\mu}}
\newcommand{\stau}{\tilde{\tau}}
\newcommand{\chp}{\tilde{\chi}^+_1}
\newcommand{\chpm}{\tilde{\chi}^\pm_1}
\newcommand{\nt}{\tilde{\chi}^0}
\newcommand{\qq}{{\mathrm q}\bar{\mathrm q}}
\newcommand{\sleppair}{\sell^+ \sell^-}
\newcommand{\nunu}{\nu \bar{\nu}}
\newcommand{\mumu}{\mu^+ \mu^-}
\newcommand{\tautau}{\tau^+ \tau^-}
\newcommand{\ellell}{\ell^+ \ell^-}
\newcommand{\nulqq}{\nu \ell {\mathrm q} \bar{\mathrm q}'}
\newcommand{\MZ}{M_{\mathrm Z}}

\newcommand {\stopm}         {\tilde{\mathrm{t}}_{1}}
\newcommand {\stops}         {\tilde{\mathrm{t}}_{2}}
\newcommand {\stopbar}       {\bar{\tilde{\mathrm{t}}}_{1}}
\newcommand {\stopx}         {\tilde{\mathrm{t}}}
\newcommand {\sneutrino}     {\tilde{\nu}}
\newcommand {\slepton}       {\tilde{\ell}}
\newcommand {\stopl}         {\tilde{\mathrm{t}}_{\mathrm L}}
\newcommand {\stopr}         {\tilde{\mathrm{t}}_{\mathrm R}}
\newcommand {\stoppair}      {\tilde{\mathrm{t}}_{1}
\bar{\tilde{\mathrm{t}}}_{1}}
\newcommand {\gluino}        {\tilde{\mathrm g}}

\newcommand {\neutralino}    {\tilde{\chi }^{0}_{1}}
\newcommand {\neutrala}      {\tilde{\chi }^{0}_{2}}
\newcommand {\neutralb}      {\tilde{\chi }^{0}_{3}}
\newcommand {\neutralc}      {\tilde{\chi }^{0}_{4}}
\newcommand {\bino}          {\tilde{\mathrm B}^{0}}
\newcommand {\wino}          {\tilde{\mathrm W}^{0}}
\newcommand {\higginoa}      {\tilde{\rm H_{1}}^{0}}
\newcommand {\higginob}      {\tilde{\mathrm H_{1}}^{0}}
\newcommand {\chargino}      {\tilde{\chi }^{\pm}_{1}}
\newcommand {\charginop}     {\tilde{\chi }^{+}_{1}}
\newcommand {\KK}            {{\mathrm K}^{0}-\bar{\mathrm K}^{0}}
\newcommand {\ff}            {{\mathrm f} \bar{\mathrm f}}
\newcommand {\bstopm} {\mbox{$\boldmath {\tilde{\mathrm{t}}_{1}} $}}
\newcommand {\Mt}            {M_{\mathrm t}}
\newcommand {\mscalar}       {m_{0}}
\newcommand {\Mgaugino}      {M_{1/2}}
\newcommand {\rs}            {\sqrt{s}}
\newcommand {\WW}            {{\mathrm W}^+{\mathrm W}^-}
\newcommand {\MGUT}          {M_{\mathrm {GUT}}}
\newcommand {\Zboson}        {${\mathrm Z}^{0}$}
\newcommand {\Wpm}           {{\mathrm W}^{\pm}}
\newcommand {\allqq}         {\sum_{q \neq t} q \bar{q}}
\newcommand {\mixang}        {\theta _{\mathrm {mix}}}
\newcommand {\thacop}        {\theta _{\mathrm {Acop}}}
\newcommand {\cosjet}        {\cos\thejet}
\newcommand {\costhr}        {\cos\thethr}
\newcommand {\djoin}         {d_{\mathrm{join}}}
\newcommand {\mstop}         {m_{\stopm}}
\newcommand {\msell}         {m_{\sell}}
\newcommand {\mchi}          {m_{\neutralino}}
\newcommand {\pp}{p \bar{p}}

\newcommand{\epair}{\mbox{${\mathrm e}^+{\mathrm e}^-$}}
\newcommand{\mupair}{\mbox{$\mu^+\mu^-$}}
\newcommand{\taupair}{\mbox{$\tau^+\tau^-$}}
\newcommand{\qpair}{\mbox{${\mathrm q}\overline{\mathrm q}$}}
\newcommand{\eeee}{\mbox{\epair\epair}}
\newcommand{\eemumu}{\mbox{\epair\mupair}}
\newcommand{\eetautau}{\mbox{\epair\taupair}}
\newcommand{\eeqq}{\mbox{\epair\qpair}}
\newcommand{\llqq}{\mbox{$\ell\ell$\qpair}}
\newcommand{\fs}{ final states}
\newcommand{\epairf}{\mbox{\epair\fs}}
\newcommand{\mupairf}{\mbox{\mupair\fs}}
\newcommand{\taupairf}{\mbox{\taupair\fs}}
\newcommand{\qpairf}{\mbox{\qpair\fs}}
\newcommand{\eeeef}{\mbox{\eeee\fs}}
\newcommand{\eemumuf}{\mbox{\eemumu\fs}}
\newcommand{\eetautauf}{\mbox{\eetautau\fs}}
\newcommand{\eeqqf}{\mbox{\eeqq\fs}}
\newcommand{\ffff}{four fermion final states}
\newcommand{\llnunu}{\mbox{$\ell^+\nu\,\ell^-\nbar$}}
\newcommand{\lnuqq}{\mbox{\lept\nubar\qpair}}
\newcommand{\zee}{\mbox{Zee}}
\newcommand{\zzg}{\mbox{ZZ/Z$\gamma$}}
\newcommand{\wenu}{\mbox{We$\nu$}}

\newcommand{\el}{\mbox{${\mathrm e}^-$}}
\newcommand{\selem}{\mbox{$\tilde{\mathrm e}^-$}}
\newcommand{\smum}{\mbox{$\tilde\mu^-$}}
\newcommand{\staum}{\mbox{$\tilde\tau^-$}}
\newcommand{\slept}{\mbox{$\tilde{\ell}^\pm$}}
\newcommand{\sleptm}{\mbox{$\tilde{\ell}^-$}}
\newcommand{\lept}{\mbox{$\ell^-$}}
\newcommand{\Hl}{\mbox{$\mathrm{L}^\pm$}}
\newcommand{\Hm}{\mbox{$\mathrm{L}^-$}}
\newcommand{\Hnu}{\mbox{$\nu_{\mathrm{L}}$}}
\newcommand{\nul}{\mbox{$\nu_\ell$}}
\newcommand{\nubar}{\mbox{$\overline{\nu}_\ell$}}
\newcommand{\nbar}{\mbox{$\overline{\nu}$}}
\newcommand{\spair}{\mbox{$\tilde{\ell}^+\tilde{\ell}^-$}}
\newcommand{\lpair}{\mbox{$\ell^+\ell^-$}}
\newcommand{\staupair}{\mbox{$\tilde{\tau}^+\tilde{\tau}^-$}}
\newcommand{\smupair}{\mbox{$\tilde{\mu}^+\tilde{\mu}^-$}}
\newcommand{\selepair}{\mbox{$\tilde{\mathrm e}^+\tilde{\mathrm e}^-$}}
\newcommand{\ch}{\mbox{$\tilde{\chi}^\pm_1$}}
\newcommand{\chpair}{\mbox{$\tilde{\chi}^+_1\tilde{\chi}^-_1$}}
\newcommand{\chm}{\mbox{$\tilde{\chi}^-_1$}}
\newcommand{\chmp}{\mbox{$\tilde{\chi}^\pm_1$}}
\newcommand{\chz}{\mbox{$\tilde{\chi}^0_1$}}
\newcommand{\dch}{\mbox{\chm$\rightarrow$\chz\lept\nubar}}
\newcommand{\dslept}{\mbox{\sleptm$\rightarrow$\lept\chz}}
\newcommand{\dH}{\mbox{$\mathrm{H}^{\pm} \rightarrow \tau^\pm \nu_\tau$}}
\newcommand{\mch}{\mbox{$m_{\tilde{\chi}^\pm_1}$}}
\newcommand{\mslept}{\mbox{$m_{\tilde{\ell}}$}}
\newcommand{\mstau}{\mbox{$m_{\staum}$}}
\newcommand{\msmu}{\mbox{$m_{\smum}$}}
\newcommand{\msele}{\mbox{$m_{\selem}$}}
\newcommand{\mchz}{\mbox{$m_{\tilde{\chi}^0_1}$}}
\newcommand{\dm}{\mbox{$\Delta m$}}
\newcommand{\dmch}{\mbox{$\Delta m_{\ch-\chz}$}}
\newcommand{\dmslept}{\mbox{$\Delta m_{\slept-\chz}$}}
\newcommand{\dmhl}{\mbox{$\Delta m_{\Hl-\Hnu}$}}
\newcommand{\w}{\mbox{W$^\pm$}}
\newcommand{\dchtwo}{\mbox{$\chpm \rightarrow \ell^\pm {\tilde{\nu}_\ell}$}}
\newcommand{\dchthree}{\mbox{$\chmp\rightarrow{\mathrm W}^\pm\chz\rightarrow\ell^\pm\nu\chz$}}
\newcommand{\chargthree}{\chpair (3-body decays)}
\newcommand{\chargtwo}{\chpair (2-body decays)}

\newcommand{\acopc}{\mbox{$\phi^{\mathrm{acop}}$}}
\newcommand{\acolc}{\mbox{$\theta^{\mathrm{acol}}$}}
\newcommand{\acop}{\mbox{$\phi_{\mathrm{acop}}$}}
\newcommand{\acol}{\mbox{$\theta_{\mathrm{acol}}$}}
\newcommand{\pz}{\mbox{$p_{\mathrm{z}}^{\mathrm{miss}}$}}
\newcommand{\ptevt}{\mbox{$p_{t}^{\mathrm{miss}}$}}
\newcommand{\ptaxic}{\mbox{$a_{t}^{\mathrm{miss}}$}}
\newcommand{\stevt}{\mbox{$p_{t}^{\mathrm{miss}}$/\Ebeam}}
\newcommand{\staxic}{\mbox{$a_{t}^{\mathrm{miss}}$/\Ebeam}}
\newcommand{\dptaxic}{\mbox{missing $p_{t}$ wrt. event axis \ptaxic}}
\newcommand{\cosevt}{\mbox{$\mid\cos\theta_{\mathrm{p}}^{\mathrm{miss}}\mid$}}
\newcommand{\axicos}{\mbox{$\mid\cos\theta_{\mathrm{a}}^{\mathrm{miss}}\mid$}}
\newcommand{\pthet}{\mbox{$\theta_{\mathrm{p}}^{\mathrm{miss}}$}}
\newcommand{\athet}{\mbox{$\theta_{\mathrm{a}}^{\mathrm{miss}}$}}
\newcommand{\dcosevt}{\mbox{$\mid\cos\theta\mid$ of missing p$_{t}$}}
\newcommand{\daxicos}{\mbox{$\mid\cos\theta\mid$ of missing p$_{t}$ wrt. event
axis}}
\newcommand{\efdsw}{\mbox{$x_{\mathrm{FDSW}}$}}
\newcommand{\acopf}{\mbox{$\Delta\phi_{\mathrm{FDSW}}$}}
\newcommand{\acopm}{\mbox{$\Delta\phi_{\mathrm{MUON}}$}}
\newcommand{\acopt}{\mbox{$\Delta\phi_{\mathrm{trk}}$}}
\newcommand{\po}{\mbox{$E_{\mathrm{isol}}^\gamma$}}
\newcommand{\qprod}{\mbox{$q1$$*$$q2$}}
\newcommand{\lcode}{lepton identification code}
\newcommand{\nctro}{\mbox{$N_{\mathrm{trk}}^{\mathrm{out}}$}}
\newcommand{\necao}{\mbox{$N_{\mathrm{ecal}}^{\mathrm{out}}$}}
\newcommand{\mout}{\mbox{$m^{\mathrm{out}}$}}
\newcommand{\nctec}{\mbox{\nctro$+$\necao}}
\newcommand{\gfract}{\mbox{$F_{\mathrm{good}}$}}
\newcommand{\zz}       {\mbox{$|z_0|$}}
\newcommand{\dz}       {\mbox{$|d_0|$}}
\newcommand{\sint}      {\mbox{$\sin\theta$}}
\newcommand{\cost}      {\mbox{$\cos\theta$}}
\newcommand{\mcost}     {\mbox{$|\cos\theta|$}}
\newcommand{\dedx}     {\mbox{$dE/dx$}}
\newcommand{\wdedx}     {\mbox{$W_{dE/dx}$}}
\newcommand{\xe}     {\mbox{$x_E$}}

\newcommand{\mH}{\mbox{$m_{\mathrm{H}^+}$}}
\newcommand{\p}     {\mbox{$\pm$}}
\newcommand{\ssix}     {\mbox{$\protect\sqrt{s}$~=~161~GeV}}
\newcommand{\sseven}     {\mbox{$\protect\sqrt{s}$~=~172~GeV}}
\newcommand{\seight}     {\mbox{$\protect\sqrt{s}$~=~183~GeV}}
\newcommand{\snine}      {\mbox{$\protect\sqrt{s}$~=~189~GeV}}
\newcommand{\sthree}     {\mbox{$\protect\sqrt{s}$~=~130--136~GeV}}
\newcommand{\mrecoil}     {\mbox{$m_{\mathrm{recoil}}$}}
\newcommand{\llmass}     {\mbox{$m_{ll}$}}
\newcommand{\sml}{\mbox{Standard Model \llnunu\ events}}
\newcommand{\sme}{\mbox{Standard Model events}}
\newcommand{\sig}{events containing a lepton pair plus missing transverse momentum}
\newcommand{\wpair}{\mbox{$\mathrm{W}^+\mathrm{W}^-$}}
\newcommand{\dW}{\mbox{W$^-\rightarrow$\lept\nubar}}
\newcommand{\dsele}{\mbox{\selem$\rightarrow$ e$^-$\chz}}
\newcommand{\eeeell}{\mbox{\epair$\rightarrow$\epair\lpair}}
\newcommand{\eell}{\mbox{\epair\lpair}}
\newcommand{\llgam}{\mbox{$\ell^+\ell^-(\gamma)$}}
\newcommand{\nunugam}{\mbox{$\nu\bar{\nu}\gamma\gamma$}}
\newcommand{\acope}{\mbox{$\Delta\phi_{\mathrm{EE}}$}}
\newcommand{\nee}{\mbox{N$_{\mathrm{EE}}$}}
\newcommand{\eesum}{\mbox{$\Sigma_{\mathrm{EE}}$}}
\newcommand{\at}{\mbox{$a_{t}$}}
\newcommand{\spp}{\mbox{$p$/\Ebeam}}
\newcommand{\acoph}{\mbox{$\Delta\phi_{\mathrm{HCAL}}$}}
\newcommand{\ACOP}{\mbox{$\phi_{\mathrm{acop}}$}}
\newcommand{\XT}{\mbox{$x_T$}}
\newcommand{\XONE}{\mbox{$x_1$}}
\newcommand{\XTWO}{\mbox{$x_2$}}
\newcommand{\MLL}{\mbox{$m_{\ell\ell}$}}
\newcommand{\MRECOIL}{\mbox{$m_{\mathrm{recoil}}$}}
\newcommand {\mm}       {\mu^+ \mu^-}
\newcommand {\emu}         {\mathrm{e}^{\pm} \mu^{\mp}}
\newcommand {\et}         {\mathrm{e}^{\pm} \tau^{\mp}}
\newcommand {\mt}         {\mu^{\pm} \tau^{\mp}}
\newcommand {\lemu}       {\ell=\mathrm{e},\mu}
\newcommand{\Zz}{\mbox{${\mathrm{Z}^0}$}}

\newcommand{\ZP}[3]    {Z. Phys. {\bf C#1} (#2) #3.}
\newcommand{\PL}[3]    {Phys. Lett. {\bf B#1} (#2) #3.}
\newcommand{\etal}     {{\it et al}.}

\newcommand{\Ecm}{\mbox{$E_{\mathrm{cm}}$}}
\newcommand{\Ebeam}{\mbox{$E_{\mathrm{beam}}$}}
\newcommand{\ipb}{\mbox{pb$^{-1}$}}
\newcommand{\wrt}{with respect to}
\newcommand{\sm}{Standard Model}
\newcommand{\smb}{Standard Model background}
\newcommand{\smp}{Standard Model processes}
\newcommand{\smc}{Standard Model Monte Carlo}
\newcommand{\mc}{Monte Carlo}
\newcommand{\btb}{back-to-back}
\newcommand{\tp}{two-photon}
\newcommand{\tpb}{two-photon background}
\newcommand{\tpp}{two-photon processes}
\newcommand{\lp}{lepton pairs}
\newcommand{\vto}{\mbox{$\tau$ veto}}

\newcommand{\gsim}{\;\raisebox{-0.9ex}
           {$\textstyle\stackrel{\textstyle >}{\sim}$}\;}
\newcommand{\lsim}{\;\raisebox{-0.9ex}{$\textstyle\stackrel{\textstyle<}
           {\sim}$}\;}

\newcommand{\degree}    {^\circ}
%
\newcommand{\roots}     {\protect\sqrt{s}}
%
%
\newcommand{\thrust}    {T}
\newcommand{\nthrust}   {\hat{n}_{\mathrm{thrust}}}
\newcommand{\thethr}    {\theta_{\,\mathrm{thrust}}}
\newcommand{\phithr}    {\phi_{\mathrm{thrust}}}
\newcommand{\acosthr}   {|\cos\thethr|}
\newcommand{\thejet}    {\theta_{\,\mathrm{jet}}}
\newcommand{\acosjet}   {|\cos\thejet|}
\newcommand{\thmiss}    { \theta_{miss} }
\newcommand{\cosmiss}   {| \cos \thmiss |}
%
%
\newcommand{\Evis}      {E_{\mathrm{vis}}}
\newcommand{\Rvis}      {E_{\mathrm{vis}}\,/\roots}
\newcommand{\Mvis}      {M_{\mathrm{vis}}}
\newcommand{\Rbal}      {R_{\mathrm{bal}}}
%
%
\newcommand{\phiacop}   {\phi_{\mathrm{acop}}}
%
%
\newcommand{\LS}      {\mbox{$L_S$}}
\newcommand{\LB}      {\mbox{$L_B$}}
\newcommand{\LR}      {\mbox{$L_R$}}
\newcommand{\PS}      {\mbox{$P_S(x_i)$}}
\newcommand{\PB}      {\mbox{$P_B(x_i)$}}
\newcommand{\signine}   {\mbox{$\sigma_{95}$}}
\newcommand{\expsig}   {\mbox{$\langle\signine\rangle$}}
\newcommand{\Lsigs}      {\mbox{$L_i(\sigma_{s})$}}
\newcommand{\Lsigsn}      {\mbox{$L_i(\sigma_{s}^{189})$}}
\newcommand{\sigs}      {\sigma_{s}}
\newcommand{\lri}      {L_{R_i}}
\newcommand{\mf}     {\mbox{$(m-15)/2$}}
\newcommand{\dstau}{\mbox{$\staum\rightarrow\tau^-\chz$}}
%
%
%
\newcommand{\PhysLett}  {Phys.~Lett.}
\newcommand{\PRL} {Phys.~Rev.\ Lett.}
\newcommand{\PhysRep}   {Phys.~Rep.}
\newcommand{\PhysRev}   {Phys.~Rev.}
\newcommand{\NPhys}  {Nucl.~Phys.}
\newcommand{\NIM} {Nucl.~Instr.\ Meth.}
\newcommand{\CPC} {Comp.~Phys.\ Comm.}
\newcommand{\ZPhys}  {Z.~Phys.}
\newcommand{\IEEENS} {IEEE Trans.\ Nucl.~Sci.}
%
%
\newcommand{\OPALColl}  {OPAL Collab.}
\newcommand{\JADEColl}  {JADE Collab.}
%
\newcommand{\onecol}[2] {\multicolumn{1}{#1}{#2}}
\newcommand{\ra}        {\rightarrow}   


\begin{titlepage}
%
%
\begin{center}{\large EUROPEAN LABORATORY FOR PARTICLE PHYSICS
}\end{center}\bigskip
\begin{flushright}
    \PRnum\\
    3$^{\mathrm{rd}}$ September 1999
\end{flushright}
\bigskip
%
%
\begin{center}
    \huge\bf\boldmath
    Search for Anomalous Production of \\ 
    Acoplanar Di-lepton Events  \\
    in \epair\ collisions \\ 
    at $\sqrt{s} = 183$ and 189 GeV
\end{center}
\bigskip\bigskip
%
%
\begin{center}
 \LARGE
The OPAL Collaboration
\bigskip
\bigskip
\bigskip
\end{center}
%
%
\begin{abstract}

A selection of di-lepton events with significant missing transverse momentum 
has been performed using a total data sample of 237.4~pb$^{-1}$
at \epair\ centre-of-mass energies of 183~GeV and 189~GeV.
The observed numbers of events --- 78 at 183 GeV and 301 at 189 GeV  ---
are consistent with the numbers expected from Standard Model processes, 
which arise predominantly from \wpair\ production with each W decaying
leptonically. 
This topology is also an experimental
signature for the pair production of new particles that decay
to a charged lepton accompanied by one or more
invisible particles.
Discrimination techniques are described that optimise 
the sensitivity to particular new physics channels.
No evidence for new phenomena is apparent and model independent limits 
are presented on the production cross-section times branching ratio squared
for sleptons and for leptonically decaying charginos and charged Higgs.
Assuming a 100\% branching ratio for the decay
$\sell^\pm_R \rightarrow  {\ell^\pm} \nt_1$, where $\nt_1$ is the
lightest neutralino, we exclude at 95\% CL:
right-handed smuons with masses below 82.3~GeV for 
\mbox{$\msmu - \mchz > 3$}~GeV and
right-handed staus with masses below 81.0~GeV for 
\mbox{$\mstau - \mchz > 8$}~GeV.
Right-handed selectrons are excluded at 95\% CL for 
masses below 87.1~GeV for \mbox{$\msele - \mchz > 5$}~GeV,
within the framework of the
Minimal Supersymmetric Standard Model assuming
$\mu < -100$~GeV and $\tan{\beta} = 1.5$.
Charged Higgs bosons, H$^{\pm}$, are excluded at 95\% CL for masses
below 82.8~GeV, assuming a 100\% branching ratio for the decay \dH .

\end{abstract}
 \bigskip
\bigskip\bigskip
\begin{center}
{\large (To be submitted to Eur.~Phys.~J.~C.)}
\end{center}

\end{titlepage}
 \begin{center}{\Large        The OPAL Collaboration
}\end{center}\bigskip
\begin{center}{
G.\thinspace Abbiendi$^{  2}$,
K.\thinspace Ackerstaff$^{  8}$,
G.\thinspace Alexander$^{ 23}$,
J.\thinspace Allison$^{ 16}$,
K.J.\thinspace Anderson$^{  9}$,
S.\thinspace Anderson$^{ 12}$,
S.\thinspace Arcelli$^{ 17}$,
S.\thinspace Asai$^{ 24}$,
S.F.\thinspace Ashby$^{  1}$,
D.\thinspace Axen$^{ 29}$
% ...
}\end{center}\bigskip
\bigskip
$^{  1}$School of Physics and Astronomy, University of Birmingham,
Birmingham B15 2TT, UK
\newline
$^{  2}$Dipartimento di Fisica dell' Universit\`a di Bologna and INFN,
I-40126 Bologna, Italy
\newline
$^{  3}$Physikalisches Institut, Universit\"at Bonn,
D-53115 Bonn, Germany
\newline
% ...
$^{  5}$Cavendish Laboratory, Cambridge CB3 0HE, UK
\newline
% ...
$^{  8}$CERN, European Organisation for Particle Physics,
CH-1211 Geneva 23, Switzerland
\newline
$^{  9}$Enrico Fermi Institute and Department of Physics,
University of Chicago, Chicago IL 60637, USA
\newline
% ...
$^{ 12}$Indiana University, Department of Physics,
Swain Hall West 117, Bloomington IN 47405, USA
\newline
% ...
$^{ 16}$Department of Physics, Schuster Laboratory, The University,
Manchester M13 9PL, UK
\newline
$^{ 17}$Department of Physics, University of Maryland,
College Park, MD 20742, USA
\newline
% ...
$^{ 23}$Department of Physics and Astronomy, Tel Aviv University,
Tel Aviv 69978, Israel
\newline
$^{ 24}$International Centre for Elementary Particle Physics and
Department of Physics, University of Tokyo, Tokyo 113-0033, and
Kobe University, Kobe 657-8501, Japan
\newline
% ...
$^{ 29}$University of British Columbia, Department of Physics,
Vancouver BC V6T 1Z1, Canada
\newline
% ...
 Debrecen, Hungary.
\newline
\clearpage

\section{Introduction}
\label{sec:intro}

We report on a set of selected events containing two oppositely
charged leptons and 
significant missing transverse momentum.
Data are analysed from e$^+$e$^-$ collisions at LEP at centre-of-mass
energies of 182.7 and 188.7~GeV
with integrated luminosities corresponding to
 56.4 \ipb\ and 181.0 \ipb , respectively.
The number of observed events and their studied properties are found to be
consistent with the expectations for  \smp , which are dominated by 
the \llnunu\ final state ($\ell$ = e, $\mu, \tau$) arising from \wpair\
production in which both
W bosons decay leptonically: $\dW$.

This topology is also an experimental
signature for the pair production of new particles that decay
to produce a charged lepton accompanied by one or more
invisible particles, such as neutrinos or 
the hypothesised lightest stable supersymmetric~\cite{SUSY} particle (LSP), 
which may be the lightest neutralino, $\nt_1$, or the gravitino, $\tilde{G}$.
Experimentally, invisible particles may also be weakly interacting neutral
particles with long lifetimes, which decay outside the
detector volume.  We present the results of 
searches for the following new particle decays:
\begin{description}
\item[charged scalar leptons (sleptons):]
$\sell^\pm \rightarrow  {\ell^\pm} \nt_1$ (or $\sell^\pm \rightarrow  {\ell^\pm} \tilde{G}$),
where $\sell^\pm$ may be a selectron ($\sele$), smuon ($\smu$) or stau
($\stau$) and $\ell^\pm$ is the corresponding charged lepton.
\item[charged Higgs:] $\mathrm{H}^{\pm} \rightarrow \tau^\pm \nu_\tau$.
\item[charginos:] $\chpm \rightarrow \ell^\pm \snu$ (``2-body'' decays)
\ \   or \ \ 
$\chpm \rightarrow \ell^\pm \nu \chz$ (``3-body'' decays).
\end{description}

The search for
charged scalar leptons provides constraints on the 
selectron mass and indirectly the 
electron-sneutrino mass (in models where these
are related). These searches are therefore also relevant to
interpreting the results of searches for chargino and neutralino
production since the production cross-section and branching 
ratios depend on the slepton masses.

In most respects the analysis is similar to our published searches
at centre-of-mass energies of 161, 172 and 183 GeV~\cite{paper172,paper183}.  
The analysis is performed in two stages.
The first stage consists of a general selection for all possible
\sig\ (Section~\ref{sec:gen}).
In this context the \sm\ \llnunu\ events are considered as signal
in addition to the possible new physics sources.
All \smp\ that do not lead to \llnunu\ final states --- e.g. \eell\ and
\llgam\ --- are considered as background and are reduced to a
rather low level by the event selection.
In the second stage the detailed properties of
the events (e.g. the type of leptons observed and their momenta), which 
vary greatly depending on the type of new particles considered and on
free parameters within the models, are used to separate as far as  possible 
the events consistent with
potential new physics sources from  \wpair\ and other \smp\
  (Section~\ref{sec:like}).

The other LEP collaborations have published searches for sleptons in 
this channel using data at $\roots\leq$183 GeV~\cite{olep}.

In this paper we describe fully only those aspects in which the second
stage of the analysis differs significantly from~\cite{paper183}.
These are:
\begin{itemize}
\item
The use of the acolinearity of the event as an additional likelihood variable,
and use of the fact that the momentum distributions employed in the likelihood 
calculation vary with acolinearity.
\item
The use of an extended maximum likelihood technique to calculate the
upper limits on the cross-section times branching ratio squared.
\end{itemize}

\section{OPAL Detector and Monte Carlo Simulation}
\label{opaldet}

A detailed description of the  OPAL detector can be found 
elsewhere~\cite{ref:OPAL-detector}.

The central detector consists of a system of chambers providing
charged particle tracking over 96\% of the full solid angle
inside a 0.435~T uniform magnetic field parallel to the beam axis. 
It consists of a two-layer
silicon micro-strip vertex detector, a high precision drift chamber,
a large volume jet chamber and a set of $z$-chambers  that measure 
the track coordinates along the beam direction. 

A lead-glass electromagnetic
calorimeter is located outside the magnet coil.  It provides, in
combination with the forward detectors, which are
lead-scintillator sandwich calorimeters and, at smaller angles,
silicon tungsten calorimeters, geometrical acceptance with excellent
hermeticity down to approximately 25~mrad.

The magnet return yoke is instrumented for hadron calorimetry 
and consists of barrel and endcap sections along with pole tip detectors that
together cover the region $|\cos \theta |<0.99$.
Outside the hadron calorimeter, four layers of muon chambers 
cover the polar angle range of $|\cos \theta |<0.98$. 
Arrays of thin scintillating tiles have been installed in the
endcap region to improve trigger performance, time resolution and hermeticity
for experiments at LEP~2~\cite{tenim}. 
Of particular relevance to this analysis are the four layers
of scintillating tiles (the MIP plug)  installed at 
each end of the OPAL detector covering
the angular range $43 < \theta < 220$~mrad.

The following \smp\ are simulated.
4-fermion production is simulated using the grc4f~\cite{grc4f}
generator at $\roots$=183~GeV, and using the {\sc Koralw}~\cite{koralw}
generator at $\roots$=189~GeV.  {\sc Koralw} uses the grc4f
matrix elements to calculate the four-fermion cross-sections including
interference effects and includes a detailed description of hard
radiation from initial, intermediate and final state charged particles.
Two-photon processes are generated using the program of
Vermaseren~\cite{vermaseren} and grc4f for \eell , and using  
{\sc Phojet}~\cite{phojet}, {\sc Herwig}~\cite{herwig} and grc4f for \eeqq .
Because of the large total cross-section for  \eeee , \eemumu\ and
\eeqq , soft cuts are applied at the generator level to preselect
events that might possibly lead to background in the selection of
\llnunu\ final states.
No generator level cuts are applied to the \eetautau\ generation.
The production of lepton pairs is generated using 
{\sc Bhwide}~\cite{bhwide} and {\sc Teegg}~\cite{teegg} for $\ee(\gamma)$, 
 and using {\sc
  Koralz}~\cite{koralz} for $\mumu(\gamma)$ and $\tautau(\gamma)$.
The production of quark pairs, \qpair(g), is generated using {\sc Pythia}~\cite{pythia}
and the final state \nunugam\ is generated using 
{\sc Nunugpv98}~\cite{NUNUGPV} and {\sc Koralz}.

Slepton pair production is generated using {\sc
  Susygen}~\cite{SUSYGEN}.
Chargino pair production is generated using {\sc Dfgt}~\cite{DFGT} for
three-body decays, and  {\sc Susygen} for two-body decays.
Charged Higgs boson pair production  is generated using
{\sc Hzha}~\cite{HZHA} and {\sc Pythia}.

All \sm\ and new physics \mc\ samples are processed with a full simulation  
of the OPAL detector \cite{gopal} and subjected to the same
reconstruction and analysis programs as used for the OPAL data.

\section{General Selection of Di-lepton Events with Significant Missing Momentum}
\label{sec:gen}

The general selection of acoplanar lepton pair events, which selects
events containing low multiplicity jets with significant missing
transverse momentum, \ptevt , is described in detail
in~\cite{paper172}.  In~\cite{paper183} we made use of the improved
hermeticity for non-showering particles in the forward direction provided 
by the MIP plug.
Subsequent modifications have been made for the analysis of the data
taken at 189~GeV, the most important of which was prompted by Monte
Carlo studies which showed that, in the majority of \smb\ events (non
$\llnunu$) accepted by
the general selection, \ptevt\ was
overestimated due to the
mis-measurement of tracks and clusters.
In the current analysis the uncertainty on \ptevt\ is calculated from
the measurement uncertainty on the observed tracks and clusters, and 
the requirement in~\cite{paper172} that  \ptevt\ exceed
certain fixed cuts is replaced by the requirement that it exceed
the cut values by at least one standard deviation of the calculated 
measurement uncertainty.
These changes -- both the MIP-plug cut and the \ptevt\ significance, 
have markedly reduced the residual backgrounds and have allowed
the selection efficiency to be substantially increased by removing 
many of the cuts of selection II which are now redundant.
In brief, the essence of event selection II is now :
\begin{itemize}
\item{\ptevt /\Ebeam\ should significantly exceed 0.05.}
\item{The scaled missing transverse momentum with respect to the 
transverse thrust axis, \staxic\, should exceed 0.022 for events
with low acoplanarity.}
\item{Events with values of \ptevt /\Ebeam\ which could potentially be 
balanced by beam energy muons in the MIP-plug acceptance are rejected
if evidence for such forward-going particles is seen in these
scintillators.}
\end{itemize}

The numbers of events passing the general selection at each centre-of-mass
energy in the data are compared to the \smc\ predictions 
in Table~\ref{tab-samples}. 
The total number of events predicted by the  \sm\ is given, 
together with a breakdown into the
contributions from individual processes.
The number of observed candidates  is consistent with the 
expectation from \sm\ sources, which is dominated by 
the \llnunu\ final state arising mostly from \wpair\ production in which both W's
decay leptonically: $\dW$.

\begin{table}[htb]
\centering
\begin{tabular}{||r||r|r||r|r|r|r|r||}
\hline \hline
$\sqrt{s}$
 (GeV)& data & SM         &  \llnunu   & \eell    & \llqq    & \llgam   & \nunugam \\
\hline \hline
  183 &   78 &  81.4\p0.8 &  77.5\p0.7 & 3.4\p0.5 & 0.07\p0.03 & 0.31\p0.04 & 0.06\p0.03 \\
  189 &  301 & 303.3\p1.9 & 292.5\p1.6 & 4.4\p0.8 & 1.3\p0.1 & 4.6\p0.4 & 0.46\p0.04 \\
\hline \hline
\end{tabular}
\caption[]{\sl
  \protect{\parbox[t]{15cm}{
Comparison between data and \mc\ of the 
number of events passing the general selection at each centre-of-mass
energy. 
The total number of events predicted by the  \sm\ is given, 
together with a breakdown into the
contributions from individual processes.
The \mc\ statistical errors are shown. 
\label{tab-samples}
}} }
\end{table}

\begin{figure}[htbp]
 \epsfxsize=\textwidth
 \epsffile{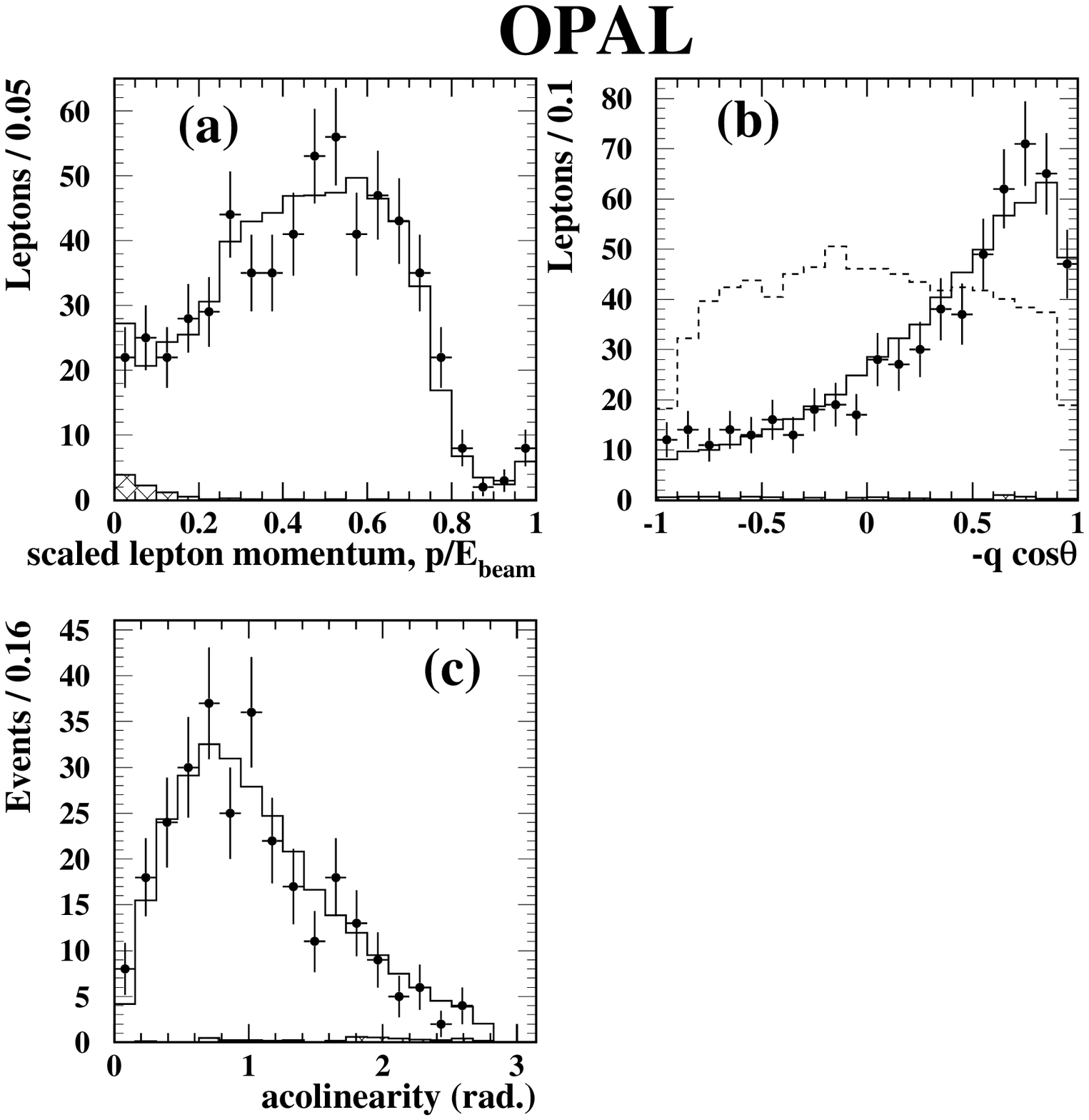}
 \caption{\sl Distributions of (a) the lepton momentum divided by the beam
   energy, (b) $-q \cos\theta$ and (c) acolinearity (in radians), for the event 
sample produced by the general selection at \snine .  
The data are shown as the points with error bars.
The \mc\ prediction for  4-fermion processes with
genuine prompt missing energy and momentum (\llnunu ) is shown as the
open histogram and the
background, arising mainly from processes with four charged leptons in
the final state, is shown as the shaded histogram.
In~(b) the dashed histogram corresponds to the distribution expected
from smuon pair production, with arbitrary normalisation.  In (a) and
(b) there are two entries per event
for events containing two identified leptons.
\label{fig-gen}
} 
\end{figure}

The second stage of the analysis, in which we distinguish between 
\sm\ and new physics sources of lepton pair
events with missing momentum, is described in Section~\ref{sec:like}.
Discrimination is provided by information on the 
lepton identification, the acolinearity of the event, and the momentum
and $-q \cos\theta$ of the observed lepton candidates, where
$q$ and $\theta$ are the charge and polar production angle of
the lepton. 
We check here on the degree to which these quantities are described by
the \smc .
The lepton identification information in the event sample produced
by the general selection at each centre-of-mass
energy is compared with the \smc\ in Table~\ref{tab-dlept}.
\begin{table}
\centering
\begin{tabular}{||c||r|r||r|r||}
\hline\hline
Lepton                           & \multicolumn{2}{c||}{$\roots$=189~GeV} & \multicolumn{2}{c||}{$\roots$=183~GeV} \\
\cline{2-5}
identification                   &  data  &  SM       &  data  &  SM        \\
\hline\hline
\epair\                          &  49    & 45.2\p0.7 &  14    & 12.1\p0.3  \\
\mupair\                         &  49    & 47.0\p0.7 &  13    & 13.7\p0.3  \\
$h^{\pm} h^{\mp}$                &  16    & 11.0\p0.5 &   1    &  2.5\p0.2  \\
$\emu$                           &  79    & 83.4\p0.9 &  20    & 25.6\p0.4  \\
$\mathrm{e}^{\pm}$ $h^{\mp}$     &  26    & 36.6\p0.6 &   8    &  9.8\p0.3  \\
$\mu^{\pm} h^{\mp}$              &  40    & 35.4\p0.6 &   8    &  9.2\p0.2  \\
$\mathrm{e}^{\pm}$, unidentified &  20    & 19.6\p0.7 &   5    &  3.8\p0.2  \\
$\mu^{\pm}$, unidentified        &  14    & 17.8\p0.5 &   7    &  3.7\p0.2  \\
$h^{\pm}$, unidentified          &   8    &  7.3\p0.3 &   2    &  0.9\p0.1  \\
\hline\hline
\end{tabular}
\caption[]{\sl
  \protect{\parbox[t]{15cm}{
The lepton identification information in the 
events passing the general selection
compared with the \smc\ at each centre-of-mass energy.
  ``$h$'' means
that the lepton is identified neither as an electron nor muon and
so is probably the product of a hadronic tau decay.
Leptonic decays of taus are usually classified
as electron or muon.  ``Unidentified'' means that only one isolated lepton has
been positively identified in the event.
}} }
\label{tab-dlept}
\end{table}
For the event sample at $\roots$=189~GeV,
Figure~\ref{fig-gen} shows the distributions 
of~(a)  the momentum scaled by the beam energy of each charged lepton candidate,
(b) the value of $-q \cos\theta$ of each charged lepton candidate, and~(c) the
acolinearity of the event.  The data  are compared with the \smc\ predictions,
which are dominated by the final state \llnunu .

The cuts used to veto two-photon background  introduce an inefficiency
in the event selection due to random detector occupancy
(principally in the SW, FD and MIP plug detectors)
that is not modelled in the \mc.
This inefficiency has been measured using randomly
triggered events collected during normal data taking.
For events with very low missing transverse momentum
the inefficiency has a value of 8.3\%
at $\sqrt{s} = 189$ GeV and 8.2\%
at $\sqrt{s} = 183$ GeV for events, and decreases
to a negligible value for events with $\stevt > 0.25$.
When quoting expected numbers of \sm\ events and selection 
efficiencies for potential new physics sources, the
variation of veto inefficiency with \ptevt\ is taken into account.

\section{Likelihood Method to Distinguish Signal and Background}
\label{sec:like}

Starting from the general selection of acoplanar di-lepton events, 
we search for the production of
new particles by using a likelihood technique which combines the information 
from a number of discriminating variables in order to distinguish between
new physics signals and \sm\ sources of such events, 
the most important of which are \llnunu\ and \eell .

Discrimination between new particle pair production and the \smb\ is 
performed by considering the likelihood that an event is consistent with 
being either signal or background.  Given an event, for which the values of 
a set of variables $x_i$ are known, 
the likelihood, \LS , of the event being consistent with the signal hypothesis
is calculated as the product of the probabilities, \PS , that the signal 
hypothesis would produce an event with variable $i$ having value $x_i$, 
$\LS = \prod_{i}\PS$.
Similarly, the likelihood of an event being consistent with the background
hypothesis is $\LB = \prod_{i}\PB$.
The discriminating quantity used is the relative likelihood, \LR , defined by:
\begin{center}
$$\LR = \frac{\LS}{\LS+\LB}.$$
\end{center}
An event with high \LR\ is signal-like and an event with low \LR\ is 
background-like.

The following quantities are used as likelihood variables ($x_i$) in the 
analysis:

(i)   Scaled momentum, $p$/\Ebeam , of each lepton,

(ii)  Acolinearity of the event, defined as the supplement of the 3-D
angle between the two leptons.

(iii) $-q \cos{\theta}$ for each lepton (smuons, staus and charged Higgs
only),

(iv)  Lepton type variable (defined in Section~\ref{sec:lept}).

Sections~\ref{sec:mom} to~\ref{sec:lept} describe the properties of each of 
these variables for signal and background.

\subsection{The Momentum Likelihood Variable}
\label{sec:mom}

The dominant \sm\ process leading to the acoplanar di-lepton signature
arises from leptonically decaying W pairs, and the momentum
distribution is highly populated between about 0.25 and 0.7
(Figure~\ref{fig-gen}(a)).  
The kinematics for the signal vary considerably
with the mass difference, \dm , between the parent particle (e.g.
selectron) and the invisible daughter particle (e.g. $\nt_1$), since this 
determines how much energy is available to the lepton.  The kinematics
also vary to a lesser extent with the mass $m$ of the parent particle,
due to Lorentz boost effects.  

A significant change to the analysis with respect to earlier publications is
the inclusion of the fact that for a given $m$ and \dm , the lepton momentum 
distribution varies according to the acolinearity of the event.
Since the parent particles (e.g. sleptons) are produced back to back, 
then if an event has high acolinearity, one of the leptons will 
typically be travelling in a direction at an angle greater than
$\pi$/2 to that of the parent particle (in the lab. frame).  In this
case, the lepton momentum in the lab. frame is reduced relative
to its value in the rest frame of the parent particle and the lepton
is therefore soft.
An event with low acolinearity will in general have both leptons 
travelling in similar directions to the parent particles and the Lorentz boost 
results in both leptons having high momenta, provided that the parent
particle mass is not close to the kinematic limit.
Since the Lorentz boost is stronger for low parent particle
mass, these effects are greater when $m$ is small.

\begin{figure}[htbp]
 \epsfxsize=\textwidth
 \epsffile{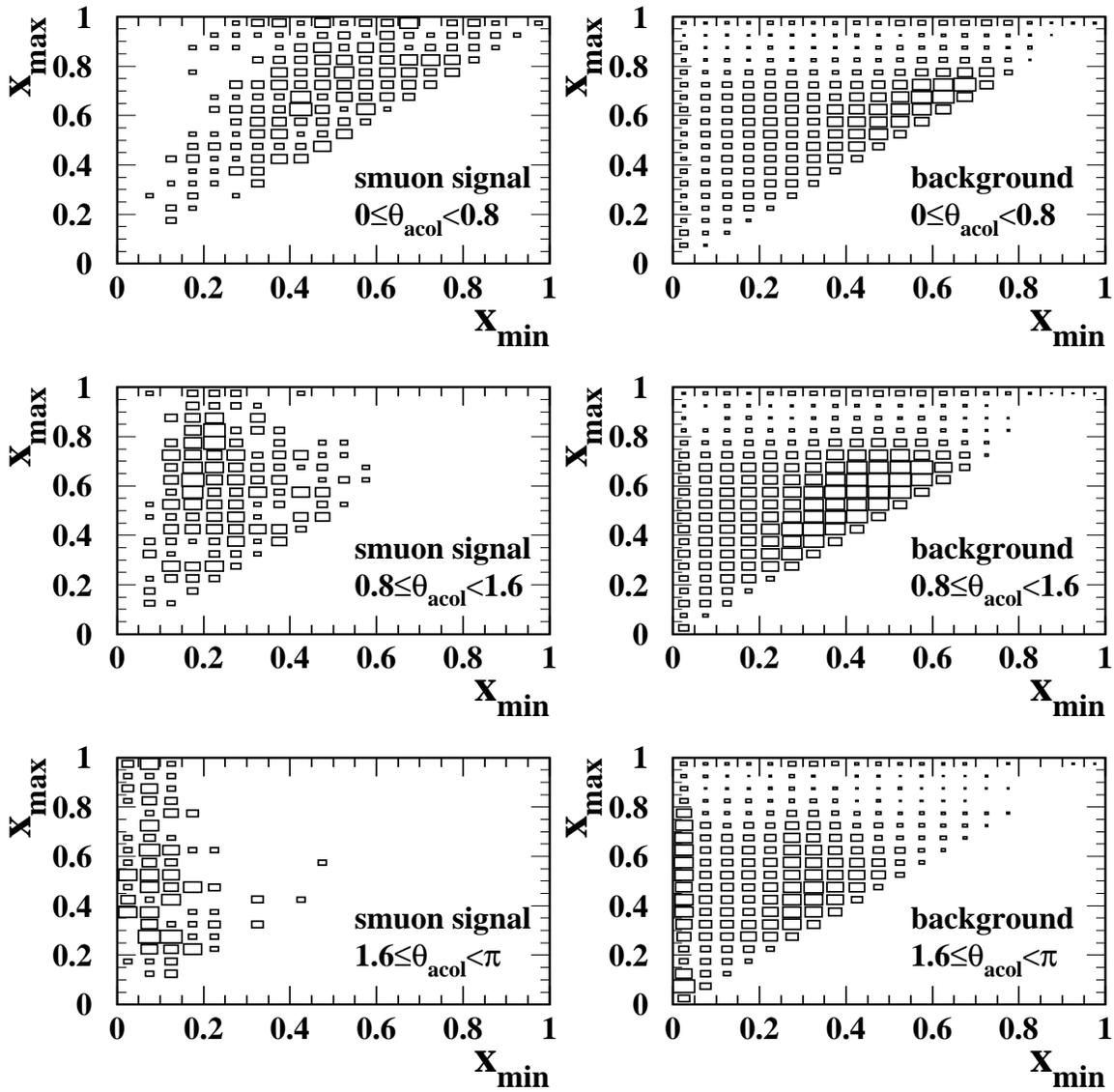}
 \caption{\sl Distributions in the ($x_{min}$,$x_{max}$) plane, where
$x_{min}$ and $x_{max}$ are the momenta of the higher and lower momentum
lepton respectively, scaled by the beam energy, for three ranges of 
acolinearity (shown in radians).  The distributions are shown for a smuon signal with 
$m$=45~GeV, \dm =45~GeV (left), and for \smc\ (right).
\label{fig-acolmom}
} 
\end{figure}

Figure~\ref{fig-acolmom} shows momentum distributions in the 
$(x_{max},x_{min})$ plane, where $x_{max}$ and $x_{min}$ are the scaled 
momenta of the higher and lower momentum leptons, respectively, for a smuon 
signal with $m$=45 GeV, \dm =45 GeV, for 3 different ranges of acolinearity.  
The corresponding plots for background are also shown.

Signal and \smc s are used to construct reference histograms for momentum 
in a grid of points in the ($m$,\dm ) plane, with $m$ ranging from 45 
to 94 GeV, and \dm\ ranging from 2 GeV to $m$.  Each of the signal momentum 
distributions and the background distribution is subdivided into the following 
3 ranges of acolinearity (in radians): $0\leq\theta_{acol}<0.8, 0.8\leq\theta_{acol}<1.6$ 
and $1.6\leq\theta_{acol}<\pi$.  Each of these distributions is then further 
subdivided according to whether the observed lepton is the higher or
lower momentum lepton.

For the searches in which the final state particles can be the decay products 
of taus (staus, charginos and charged Higgs), each momentum probability 
distribution must be further subdivided, 
as the momentum spectrum depends on the lepton identification.  
One momentum probability distribution is constructed for the case in
which the observed lepton is identified as e or $\mu$, and another for
the case in which the observed lepton is identified as a hadronically decaying
tau, or is unidentified.

Probabilities \PS\ and \PB\ for the likelihood calculation are found by reading 
from the appropriate reference histograms.

\subsection{The Acolinearity Likelihood Variable}
\label{sec:acol}

For signal, the distribution of the acolinearity angle varies with $m$ and \dm .
For low parent particle mass, the Lorentz boost results in a
tendency for the leptons to be in the directions of the parent particles, resulting in 
the acolinearity being peaked towards low values, whereas for high mass, the parent 
particles are produced close to being at rest, and the leptons have no preferred 
direction.  For background, the acolinearity distribution is peaked
towards low values due to the spin structure of the weak couplings.

\begin{figure}[htbp]
 \epsfxsize=\textwidth 
 \epsffile{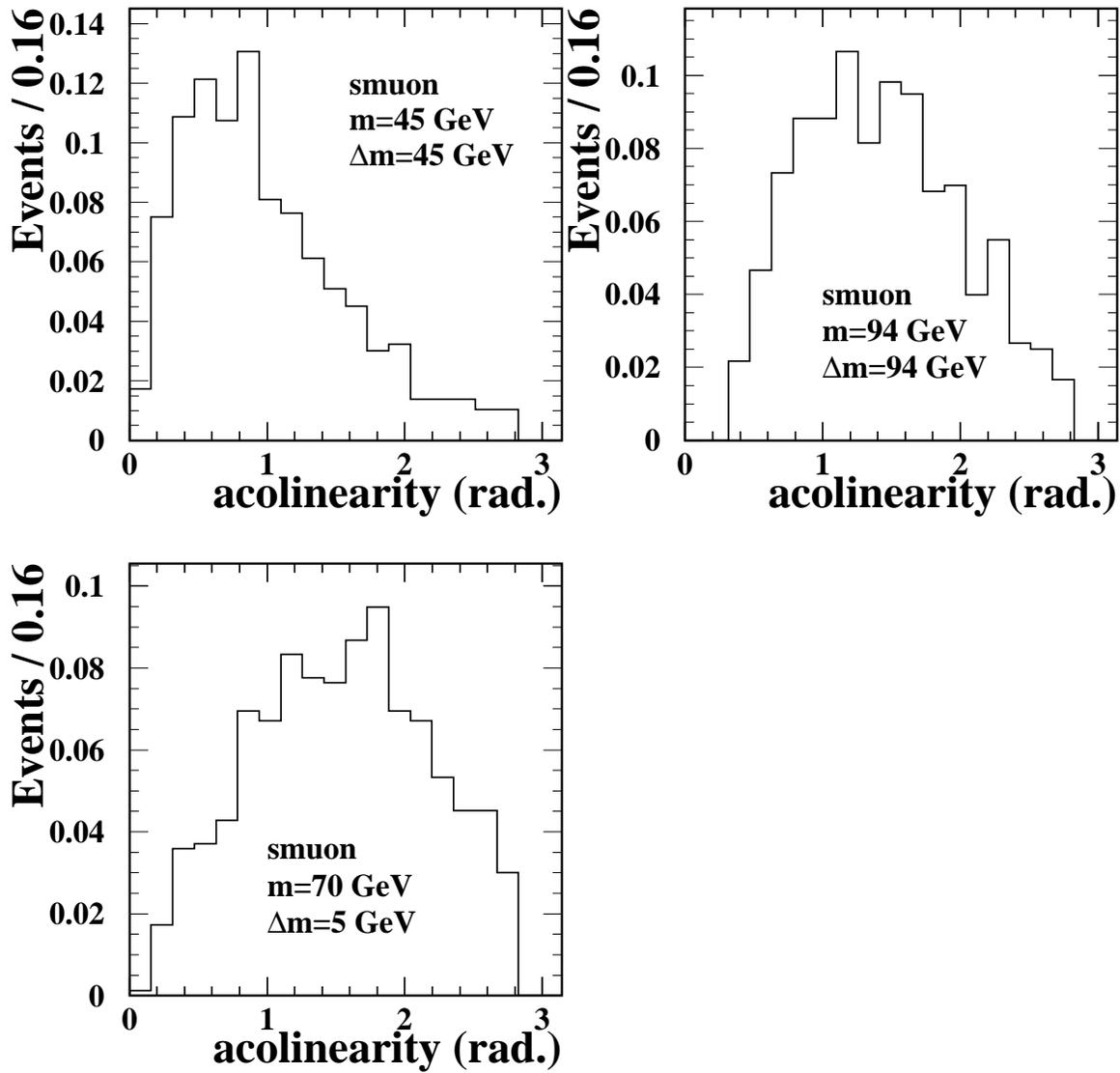}
 \caption{\sl
Distributions of acolinearity (in radians) for three example smuon signals.
\label{fig-acol}
} 
\end{figure}

Figure~\ref{fig-acol} shows some example acolinearity distributions for signal, 
which can be compared to the distributions for background and data shown in 
Figure~\ref{fig-gen}.

The use of the acolinearity as a likelihood variable is complementary to its use
in defining the momentum probability, in that the masses for
which it offers the greatest discrimination as a likelihood variable are the masses
where the gain in distinguishing power described in Section~\ref{sec:mom} is small,
and vice versa.

\subsection{The $-q \cos{\theta}$ Likelihood Variable}
\label{sec:qcos}

As described in~\cite{paper183}, the distribution of the quantity 
$-q \cos{\theta}$ , where $q$ and $\theta$ are the charge and production angle of an 
observed lepton, is forward peaked for \wpair\ production due to the dominance of 
the neutrino exchange amplitude and the V-A nature of W decay, whereas for smuon, stau 
and charged Higgs production the distribution is symmetric and peaked towards 
$|\cos{\theta}|=0$, due to the scalar nature of these particles.

This variable is not used in the likelihood calculation for selectrons or charginos
because these particles can be produced 
via $t$-channel neutralino exchange and sneutrino exchange, respectively, in 
addition to $s$-channel production.  This results in the expected $- q \cos{\theta}$ 
distribution of selectrons and charginos being model-dependent and potentially
similar to that of the \wpair\ background.

\subsection{The Lepton Type Likelihood Variable}
\label{sec:lept}

A value is assigned to the lepton type variable according to which
types of lepton are identified in the event.  There are nine possible
values, corresponding to the nine event types listed in table~\ref{tab-dlept}.

For the selectron (smuon) analysis a cut is applied 
at the same time as the general selection, requiring at least one
electron (muon) and no muons (electrons), reducing the background by
about 2/3 (depending on slepton type and $\roots$) with negligible
loss of efficiency.  With this cut applied, there are only three
possible values of the lepton type variable.

\subsection{\LR\ and \LB\ Distributions}
\label{sec:lb}

For each search channel, reference histograms are constructed for each
of the likelihood variables at each point in $m$ and \dm\ for which
signal \mc\ has been generated.  A smoothing algorithm~\cite{smooth}
is applied to the histograms to reduce the effects of statistical
fluctuations.  The reference histograms are then used to construct
\LR\ distributions.

\begin{figure}[htbp]
 \epsfxsize=\textwidth 
 \epsffile{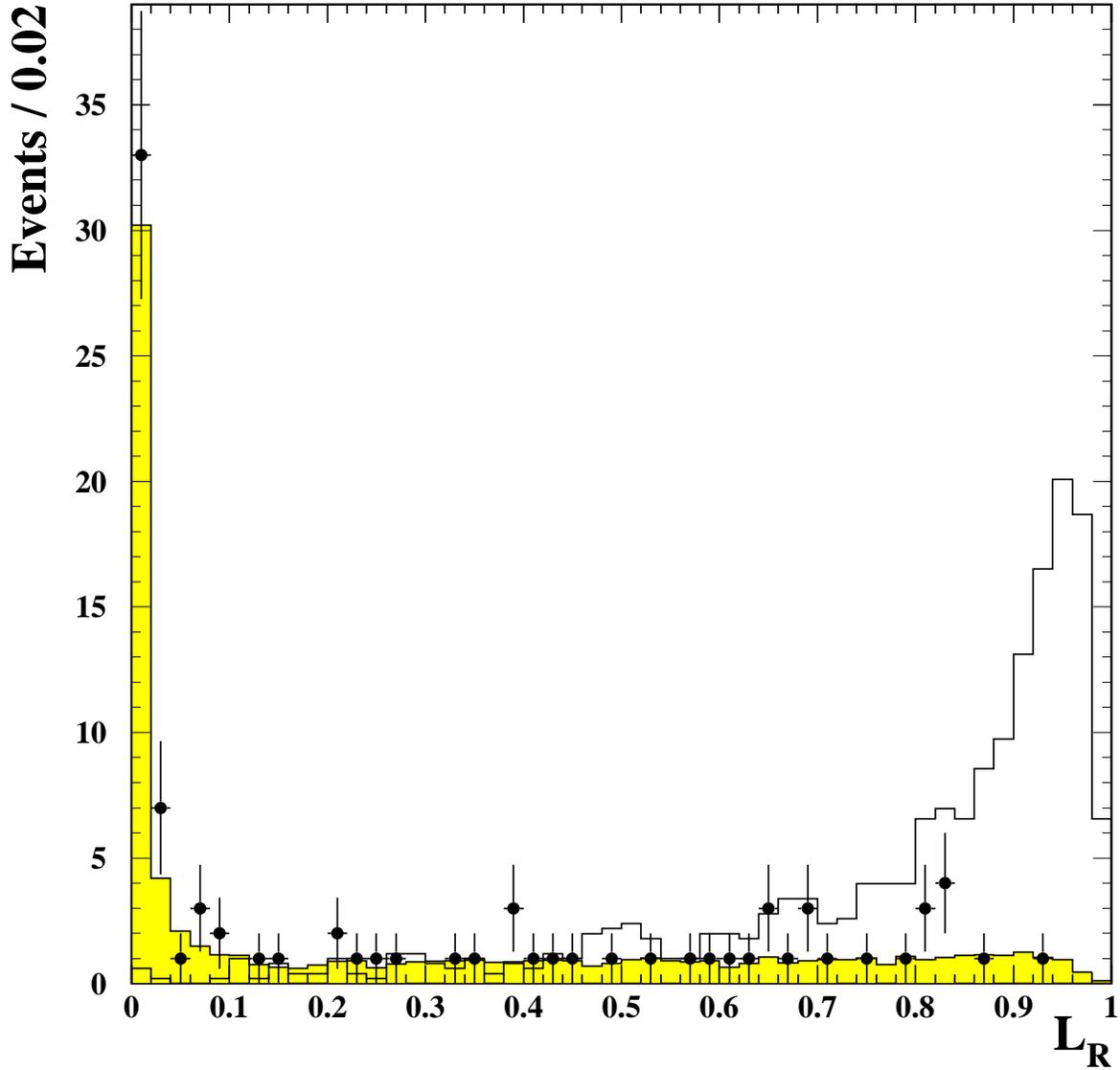}
 \caption{\sl Distributions of the relative likelihood, \LR , for
\smc\ (shaded histogram), signal (open
histogram) and data (points with error bars), in the analysis for smuons
with a mass of 80~GeV for a smuon-neutralino mass difference of 60~GeV.
\label{fig:lr}
} 
\end{figure}

\LR\ distributions for signal \mc , \smc\ and data are shown
in Figure~\ref{fig:lr} for the specific example of the analysis for
smuons with a mass of 80~GeV and a smuon-neutralino mass difference 
of 60~GeV.  There is considerable variation in the shapes of these
distributions with $m$ and \dm .

\begin{figure}[htbp]
 \epsfxsize=\textwidth 
 \epsffile{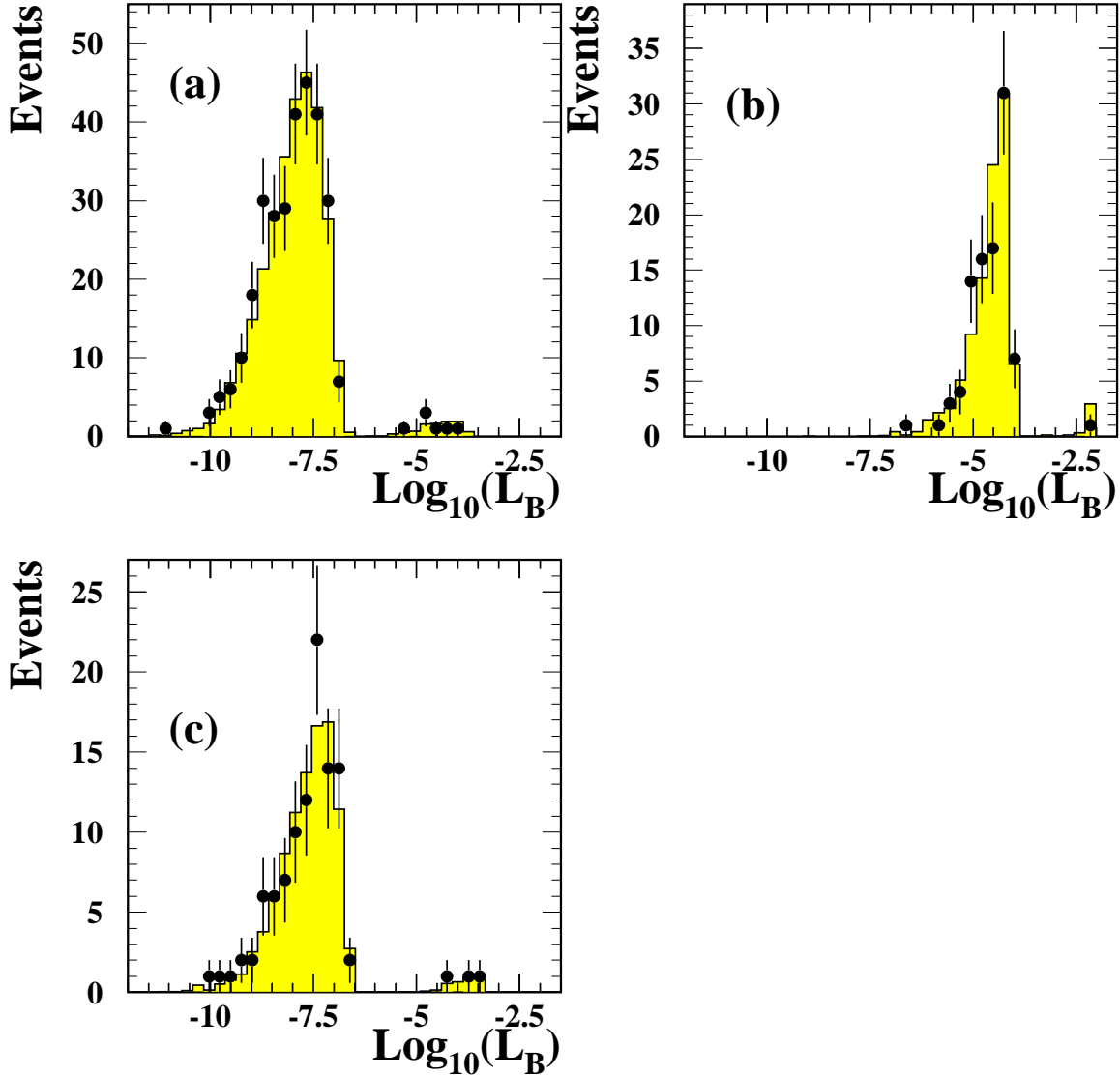}
 \caption{\sl Distributions of the background likelihood, \LB , for
\smc\ (shaded histogram) and data 
(points with error bars) for events passing the general selection, using all
the likelihood variables (a).  (b) and (c) show the same information 
after making the initial lepton identification requirements given in 
Section~\ref{sec:lept} for the selectron and smuon searches respectively.
In (b), only the variables used in the selectron analysis
are used.
\label{fig:lb}
} 
\end{figure}

A check of consistency between data and the \sm\ can be performed without 
reference
to a particular signal by comparing the \LB\ distributions for data and \sm .
Figure~\ref{fig:lb}(a) shows the \LB\ distributions for the \sm\ (histogram)
and data (points) for events passing the general selection.  All the likelihood
variables are used.  Figures~\ref{fig:lb}(b) and (c) show the same information 
after making the initial lepton identification requirements given in 
Section~\ref{sec:lept} for the selectron and smuon searches respectively.
In Figure~\ref{fig:lb}(b), only the variables used in the selectron analysis
are used.  In each of the plots, the secondary peak at high \LB\ corresponds
to events which have only one identified lepton and therefore fewer variables
entering the likelihood~\footnote{This effect cancels when the
likelihood ratio \LR\ is calculated because the events will have high
\LS\ for the same reason.}.  In all three plots, the data is in good
agreement with the \sm .

\section{Calculation of Cross-Section Limits}
\label{sec:limits}

\subsection{Introduction}
\label{sec:s95intro}

In~\cite{paper183} , the limit on the cross-section was calculated by finding
an optimised cut on the value of \LR\ as a function of $m$ and \dm\ for each
centre of mass energy, and applying
this cut to signal \mc , \smc\ and data.  The resulting efficiencies,
expected backgrounds and numbers of candidates were used to calculate 
cross-section limits using the likelihood ratio method~\cite{LR} to
combine the information.

In this paper, we describe the use of an extended maximum likelihood calculation
to determine the cross-section limits.  In this method, no cut is applied on
\LR .  Information contained in the \LR\ values of each individual candidate event, 
and in the shapes of the \LR\ distributions for signal and background are used
as input to the limit calculation, rather than the numbers of events passing a cut.
In this way, considerably more of the available information is used.

The advantage of a cut free method can be seen by considering, for
example, a case where there
is an excess of candidates passing a cut on \LR .  The information about whether
the events all lie close to the cut, or whether they are clustered towards \LR=1 
(suggesting the presence of a signal) is not used.  The use of the additional
information makes the analysis more sensitive for discovery, and at the same time
is able to set more stringent limits in the absence of signal.  The
expected sensitivity (ie., the expected upper limit on the cross-section)
is improved by as much as 20\%, depending on $m$ and \dm , using this technique.

\subsection{Extended Maximum Likelihood Technique}
\label{sec:EML}

The upper limit on the cross-section times branching ratio squared at 95\%
confidence level, \signine , is calculated by forming a likelihood, $L(\sigs )$, of the set of
\LR\ values for the data being consistent with the expected \LR\ distribution
for \sm\ plus a signal produced with cross-section times branching ratio squared 
$\sigs$.  \signine\ is the value of $\sigs$ below which 95\% 
of the area under the likelihood function lies.

\subsubsection{The Likelihood Function}
\label{sec:likefun}

Extended maximum likelihood combines standard maximum likelihood with the
Poisson probability of observing $N$ candidate events when $\nu$ are expected:

$$L = \frac{e^{-\nu}\nu^N}{N!} \prod_{i=1}^{N} P(\lri;B,S),$$

where $P(\lri;B,S)$ is the probability of event $i$ having $\LR=\lri$, given
\LR\ distributions $B$ and $S$ for background and signal.

Dropping the constant $N!$, this can be re-written:

$$\ln L = - \nu + \sum_{i=1}^{N} \ln [Q(\lri;B,S)]$$

where $Q$ is identical to $P$ but normalised to $\nu$ instead of 1
($Q=\nu P$).

The expected number of candidates $\nu$ is given by:

$$\nu = \mu_B + \epsilon  {\cal L} \omega \sigs,$$

where $\mu_B$ is the expected number of \sm\ events with non-zero \LR\ 
passing the general selection (similarly, $N$ is the number of data 
candidates with $\LR \neq 0$), $\epsilon$ is the signal selection 
efficiency of the general selection, ${\cal L}$ is the experimental 
luminosity and $\omega$ is a weight factor which takes
into account that the expected production cross-section varies with $\roots$,
but the limit on the observed cross-section is quoted at \snine.
$$ \omega_i = \frac{\sigma_i}{\sigma_{189}},$$
where $\sigma_{189}$ is the expected cross-section for \snine\
and $\sigma_i$ is the expected cross-section for the $i$'th value of
$\roots$.  
For scalar particles, for example sleptons, 
we assume that the expected cross-section
varies as $\beta^3/s$.
For spin $\frac{1}{2}$ particles, for example charginos, 
we assume that the expected cross-section
varies as $\beta/s$. 

The function $Q$ is the probability of event $i$ having $L_R=\lri$, given
\LR\ distributions $B$ and $S$ for background and signal, normalised to $\nu$.
This is given by:

$$Q = \mu_B B(\lri) + \epsilon  {\cal L} \omega \sigs S(\lri),$$

where the functions $B$ and $S$, formed using background and signal \mc\ 
respectively, are normalised to 1.

Hence the likelihood function is given by:

$$\ln L(\sigs) = -(\mu_B + \epsilon {\cal L} \omega \sigs ) + \sum_{i=1}^{N} \ln[\mu_B B(\lri) + \epsilon  {\cal L} \omega \sigs S(\lri)].$$

\subsubsection{Limit Calculation}
\label{sec:limit}

For data at a single centre-of-mass energy, the upper limit on the cross-section 
at 95\% 
confidence level is the value of \signine\ which satisfies:

$$ 0.95 = \frac{\int_0^{\signine} L(\sigs ) \mathrm{d}\sigs}{\int_0^{\infty} L(\sigs ) \mathrm{d}\sigs}.$$

The generalisation to $N_{ECM}$ values of $\roots$ is:

$$ 0.95 = \frac{\int_0^{\sigma_{95}^{189}} \prod_{i=1}^{N_{ECM}} \Lsigsn \mathrm{d}\sigs^{189}}{\int_0^{\infty} \prod_{i=1}^{N_{ECM}} \Lsigsn \mathrm{d}\sigs^{189}}.$$

where $\sigma_{s}^{189}$ is the cross-section at \snine .

\subsubsection{Verification of the Method}
\label{sec:ver}

The technique used to calculate limits was tested using a toy \mc\ to simulate
data sets for an ensemble of simulated experiments in which a signal is present
with cross-section $\sigs$.  The total number of candidates was
drawn from a Poisson distribution with mean $\nu$ and for each candidate, 
a value of \LR\ was
assigned, chosen randomly according to the sum of the expected \LR\ 
distributions for background and signal with cross-section $\sigs$.

\signine\ was calculated for 500 simulated experiments.  This was done using 
smuon signals at all $m$ and \dm\ for which \mc\ has been generated, 
for 189~GeV separately and for 183~GeV and 189~GeV combined, and for a number 
of values of $\sigs$.  In all cases,
95\% of the 500 \signine\ values were found to be greater than the
true cross-section $\sigs$,
to within the statistical error expected from the finite number of 
simulated experiments.

As a further test, the best estimate of the signal cross-section, $\sigma_{best}$,
was calculated for each simulated experiment.  This is the value of $\sigs$ at 
which the likelihood function $\Lsigs$ peaks.  The $\sigma_{best}$ distribution
from the 500 simulated experiments was in all cases found to peak at the true
cross-section.

\subsection{Limit Calculation at an Arbitrary Point in $m$ and \dm}
\label{sec:interp}

Monte Carlo signal events are available only at certain particular
values of $m$ and \dm .  The values of $m$ range typically from 
$m = 45$~GeV~\footnote{Particle masses less than 45~GeV are not considered
because these masses were accessible at LEP1 and because radiative return to 
the Z means that the event topology can be different.}
up to $m \approx \Ebeam$ in 5~GeV steps.  The values of \dm\ vary between 2 
and \dm = $m$.  In order to calculate \signine\ at an intermediate point in 
$m$ and \dm , it is necessary to be able to calculate \LR\ at that point for 
a given event, which requires the existence of reference histograms for the 
likelihood variables for any $m$ and \dm .

An algorithm has been developed to construct the reference
histograms at any intermediate value of $m$ and \dm , given the histograms at the 
four nearest signal \mc\ grid points, assuming a linear variation in the shape of 
the histograms with $m$ and \dm .  This procedure has been tested by re-constructing
histograms at gridpoints using the histograms at adjacent gridpoints.
  
For a given point $m$ and \dm , the signal and background 
\LR\ distributions and the data \LR\ values are calculated using the
interpolated reference histograms.  Signal \mc\ 
at an intermediate point is simulated using signal \mc\ events at the nearest grid point.  
This is done by re-defining the value of each likelihood 
variable for an event, such that the fraction of the corresponding reference histogram 
at the intermediate point which lies below the re-defined value is the same 
as the fraction of the histogram at the grid point which lies below the original value.

The effects of statistical fluctuations in the \LR\ distributions are reduced using
the same smoothing algorithm as applied to the reference histograms
(section~\ref{sec:lb}).

The remaining input to the limit calculation is the signal efficiency, obtained at 
intermediate values of $m$ and \dm\  by linear 2-dimensional interpolation.

\section{New Particle Search Results}
\label{results}

We present limits on the pair production
of charged scalar leptons, leptonically decaying charged Higgs bosons and 
charginos that decay to produce a charged lepton and invisible particles.

The 95\% CL upper limit on  new particle production at \snine , 
obtained by combining the data at \snine\ and \seight\ is calculated at each
kinematically allowed point on a 0.5~GeV by 0.5~GeV grid of $m$ and \dm , using
the \LR\ distributions for signal and background, the \LR\ values of the data events,
and the efficiency of the general selection at that point as input.

In addition to the \mc\ statistical error on the signal efficiency, we 
assign a 10\% 
systematic error on the estimated selection efficiency to take into
account uncertainties in trigger efficiency, detector
occupancy, lepton identification efficiency, luminosity measurement, 
the interpolation procedure, and
deficiencies in the \mc\ generators and the detector simulation.
An additional systematic error in the stau analysis is the effect of tau 
polarisation in the modelling of the stau signal.  It is possible for
the tau produced in stau decay to have any polarisation value in the
range [-1,1]~\cite{Nojiri}.  This was studied by 
using stau \mc\ events with tau polarisations of +1, 0 and $-$1 to determine 
the amount by which the expected limit 
on the cross-section times branching ratio squared is overestimated
or underestimated if a polarisation of zero is assumed when the true 
polarisation is +1 or $-$1.  The size of this effect was found to vary with 
$m$ and \dm , but to be 
always less than about 5\% , and so is included in the 10\% systematic error.

At high values of \dm\ the dominant background to 
the searches for new physics results from \wpair\ production.
High statistics \mc\ samples for this process are available that describe well
the OPAL data~\cite{wwpaper}.
In addition to the \mc\ statistical error, we assign a 10\% 
systematic error on the estimated background to take into account uncertainties
in the shapes of the \LR\ distributions and reference histograms, and in the 
interpolation procedure, and deficiencies in the \mc\ detector simulation.
At low values of \dm\ the dominant background results from \eell\
events.
The background uncertainty at low \dm\ is dominated by the limited \mc\
statistics; the
uncertainty is typically 20--80\% at low \dm . 
In setting limits the \mc\ statistical errors and other
systematics are taken into account according to the 
method described in~\cite{cousins}.

\subsection{Limits on Production Cross-section Times Branching Ratio Squared}
\label{sec:crosssec}

\begin{figure}[htbp]
 \epsfxsize=\textwidth 
 \epsffile[0 0 580 600]{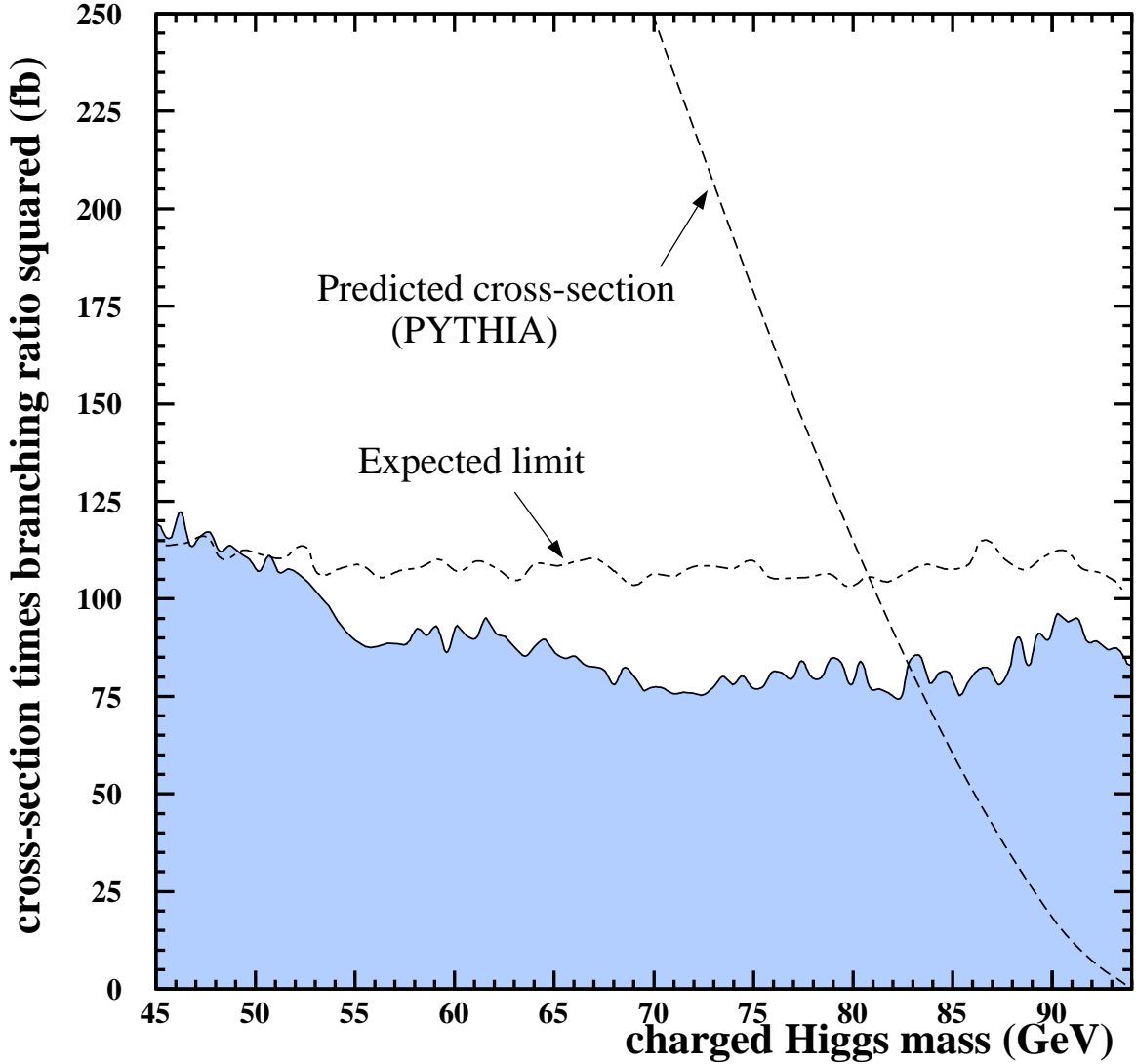}
 \caption{\sl 
The solid line shows the 95\% CL upper limit on the 
charged Higgs pair production 
cross-section times branching ratio squared for the decay \dH\
at $\protect\sqrt{s}$~=~189~GeV.
The limit is obtained by combining the 183 and 189~GeV data-sets 
assuming the \mH\ and $\protect\sqrt{s}$ 
dependence of the cross-section predicted
by {\sc Pythia}.
For comparison, the dashed curve shows the prediction from {\sc Pythia}
 at $\protect\sqrt{s}$~=~189~GeV
assuming a 100\% branching ratio for the decay \dH .
The expected limit calculated from \mc\ alone is indicated by
the dash-dotted line.
} 
\label{fig:limit_5}
\end{figure}

Limits on the production cross-section times branching ratio squared for new
physics processes are now presented in a manner intended to 
minimise the number of model assumptions.
The 95\% CL upper limits at \snine\ shown in 
Figures~\protect\ref{fig:limit_1}~--~\protect\ref{fig:limit_5} are
obtained by combining the data at the two centre-of-mass 
energies 183 and 189~GeV using
the assumption that the
cross-section varies as $\beta^3/s$ for sleptons
and $\beta/s$ for charginos.
The chosen functional forms are used for simplicity in presenting the
data and represent an approximation, particularly
for processes in which $t$-channel exchange may be important, that is,
selectron pair and chargino pair production.
In these cases the cross-section
dependence on centre-of-mass energy 
is model dependent, depending on the mass of the exchanged particles 
and the couplings of the neutralinos and charginos.  The selectron
\mc\ events were generated at $\mu = -200$~GeV and
$\tan{\beta} = 1.5$ using {\sc Susygen}.
We have found by varying $\mu$ and $\tan{\beta}$ that the above choice
gives a conservative estimate of the selection efficiency for selectrons. 

\begin{figure}[htbp]
 \epsfxsize=\textwidth 
 \epsffile[0 0 580 600]{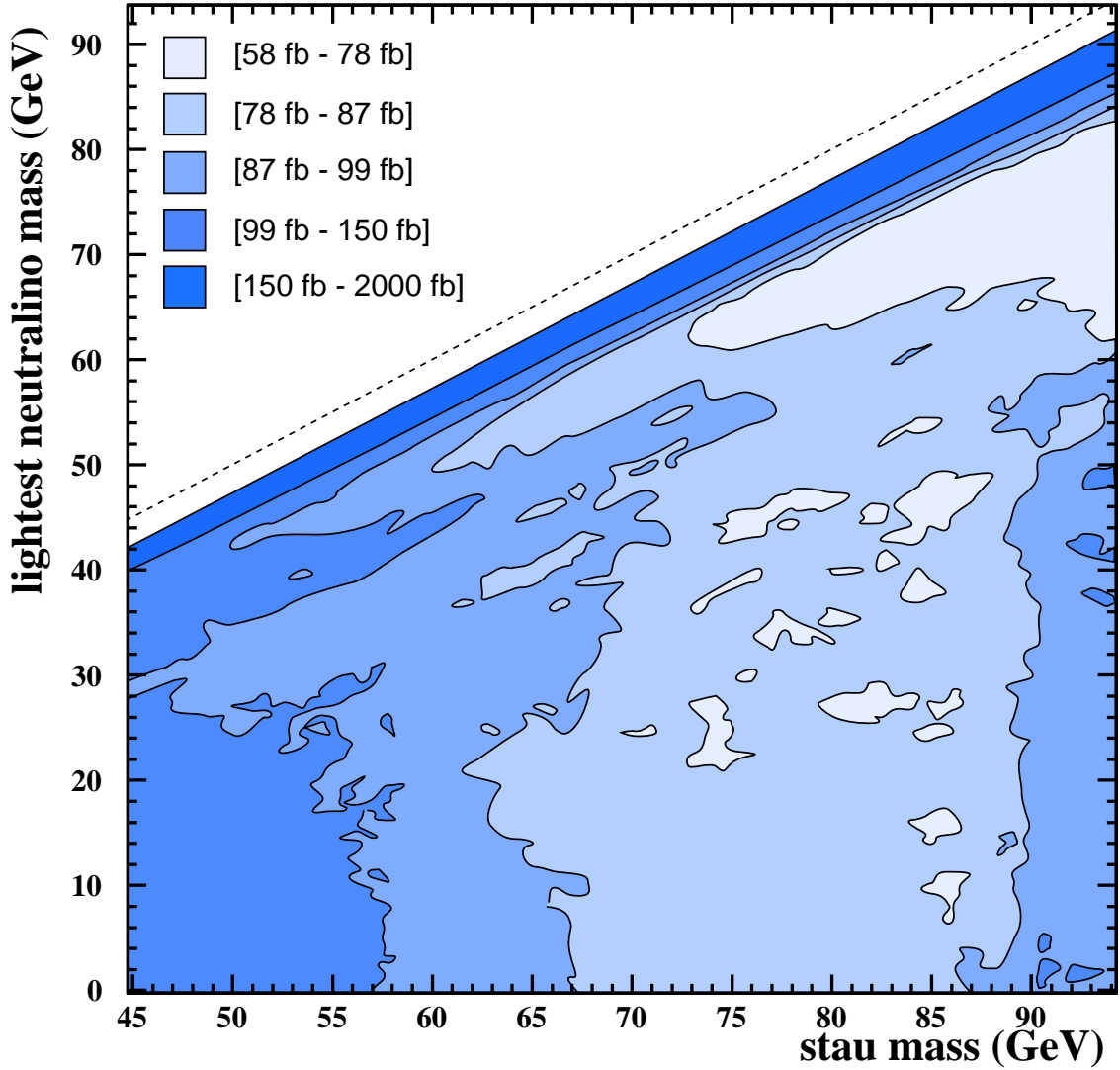}
 \caption{\sl
Contours of the 95\% CL upper limits on the stau pair
cross-section times $BR^2(\stau \rightarrow \tau \nt_1)$
at 189 GeV
based on combining the 183 and 189 GeV data-sets 
assuming a $\beta^3/s$ dependence of the cross-section.
The kinematically allowed region is indicated by the dashed line.  The
unshaded region at very low \dm\ is experimentally inaccessible in
this search.
} 
\label{fig:limit_3}
\end{figure}

Upper limits at  95\% CL on the selectron pair cross-section 
at~\snine\ times 
branching ratio squared for the decay \dsele\
are shown in Figure~\ref{fig:limit_1} as a function of selectron mass 
and lightest neutralino mass.
These limits are applicable to
$\tilde{\mathrm e}^+_{\mathrm L}\tilde{\mathrm e}^-_{\mathrm L}$ and 
$\tilde{\mathrm e}^+_{\mathrm R}\tilde{\mathrm e}^-_{\mathrm R}$ production.
The corresponding plots for  the smuon and stau pair searches are shown in  
Figures~\ref{fig:limit_2} and~\ref{fig:limit_3}, respectively.
Note that if the  LSP is the gravitino, $\tilde{G}$ (effectively massless), then for 
prompt slepton decays to a lepton and a gravitino the experimental signature 
would be the same as that
for  \dslept\ with a massless neutralino.
In this case the limits given in 
Figures~\protect\ref{fig:limit_1}~--~\protect\ref{fig:limit_3}
for $m_{\tilde{\chi}^0_1} = 0$ may be interpreted as limits on the
decay $\tilde{\ell}^-\rightarrow \ell^-\tilde{G}$

\begin{figure}[htbp]
 \epsfxsize=\textwidth 
 \epsffile[0 0 580 600]{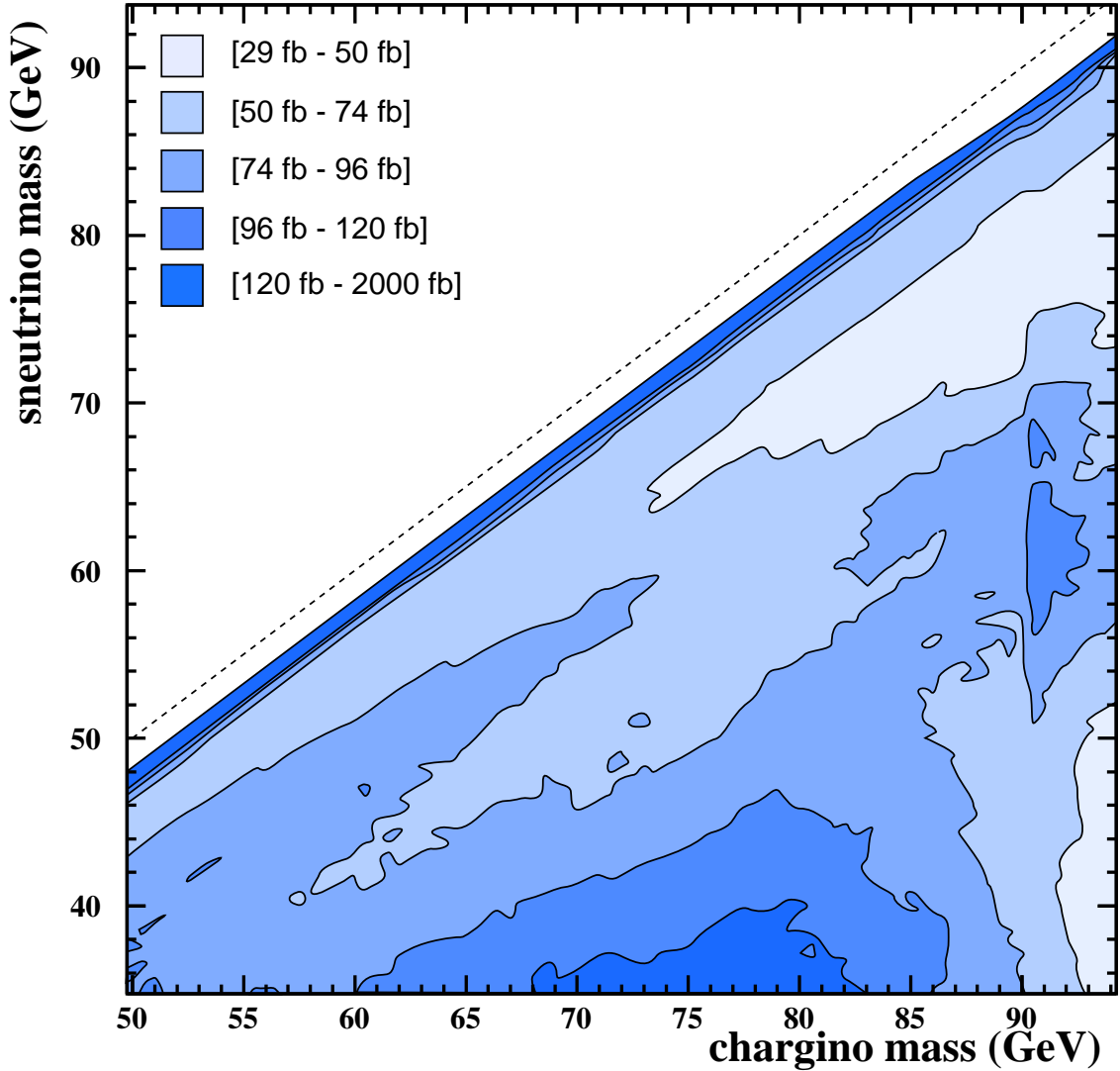}
 \caption{\sl 
Contours of the 95\% CL upper limits on the chargino pair
cross-section times branching ratio squared for 
$\chpm \rightarrow \ell^\pm \snu$ (2-body decay)
at $\protect\sqrt{s}$~=~189~GeV.
The  limits have been calculated for the 
case where the three sneutrino 
generations are mass degenerate.
The limit is obtained by combining the 183 and 189~GeV data-sets 
assuming a $\beta/s$ dependence of the cross-section.
The kinematically allowed region is indicated by the dashed line.  The
unshaded region at very low \dm\ is experimentally inaccessible in
this search.
} 
\label{fig:limit_8}
\end{figure}

\begin{figure}[htbp]
 \epsfxsize=\textwidth 
 \epsffile[0 0 580 600]{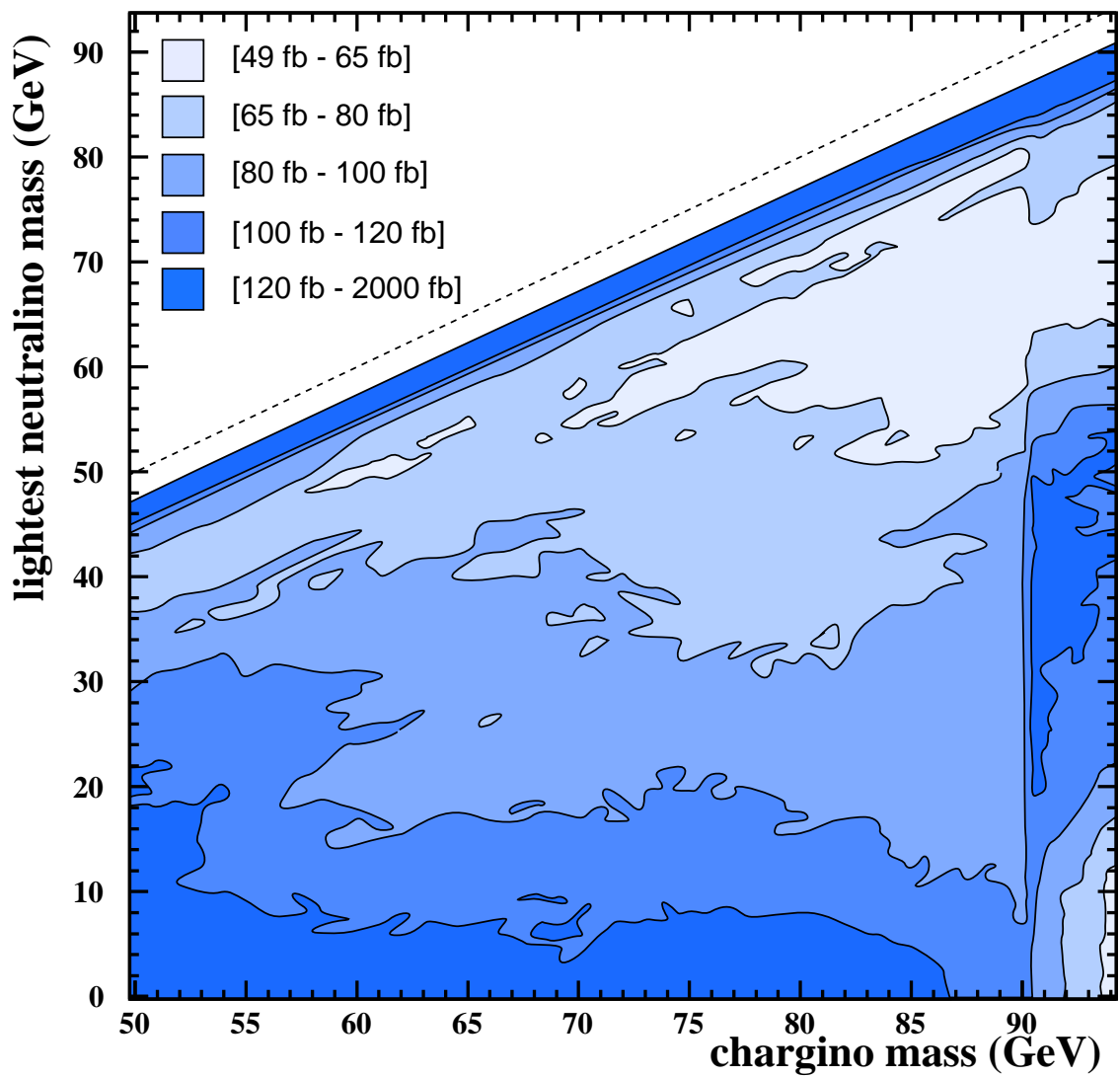}
 \caption{\sl 
Contours of the 95\% CL upper limits on the chargino pair
cross-section times branching ratio squared for 
$\chpm \rightarrow \ell^\pm \nu \chz$
 (3-body decay) at $\protect\sqrt{s}$~=~189~GeV.
The limit is obtained by combining the 183 and 189~GeV data-sets 
assuming a $\beta/s$ dependence of the cross-section.
The kinematically allowed region is indicated by the dashed line.  The
unshaded region at very low \dm\ is experimentally inaccessible in
this search.
} 
\label{fig:limit_4}
\end{figure}

The upper limit at 95\% CL on the chargino pair production 
cross-section times branching
ratio squared for the decay \dchtwo\  (2-body decay)
is shown in Figure \ref{fig:limit_8}. 
The  limit has been calculated for the 
case where the three sneutrino 
generations are mass degenerate.
The upper limit at 95\% CL on the chargino pair production 
cross-section times branching
ratio squared for the decay \dchthree\ (3-body decay)
is shown in Figure \ref{fig:limit_4}. 

The upper limit at 95\% CL on the charged Higgs boson pair production 
cross-section times branching ratio squared for the decay \dH\
is shown as a function of \mH\ as the solid line in Figure \ref{fig:limit_5}. 
The limit is obtained by combining the 183 and 189~GeV data-sets 
assuming the \mH\ and $\sqrt{s}$ 
dependence of the cross-section predicted
by {\sc Pythia}, which takes into account the effect of 
initial state radiation.
The dashed  line in Figure \ref{fig:limit_5} 
shows the prediction from {\sc Pythia}
 at $\protect\sqrt{s}$~=~189~GeV
for a 100\% branching ratio for the decay \dH .
With this assumption we set a lower limit  at 95\% CL on \mH\ of 82.8~GeV.

\subsection{Expected Limits and Confidence Levels for Consistency with Expectation}
\label{sec:expect}

Table~\ref{tab-CL1} gives the values of the following quantities for a number
of values of $m$ and \dm\ in the search for selectrons:
\begin{enumerate}

\item
The signal selection efficiency of the general selection at 189~GeV
(the efficiencies at 183~GeV are similar).

\item
The 95\% CL upper limit on the 
cross-section times branching ratio squared at 189~GeV, obtained by
combining the data at \snine\ and \seight .

\item
The expected 95\% CL upper limit on the cross-section times branching 
ratio squared in the absence of signal \expsig .
This is calculated using an ensemble of 1000 toy \mc\ experiments to 
simulate the data, in which the total number of candidates for each 
experiment is
drawn from a Poisson distribution with mean equal to the number of events 
expected from the \sm , and for each candidate, 
a value of $\LR$ is assigned, chosen randomly according to the 
expected $\LR$ distribution for \smp .
The expected limit at a given point in $m$ and \dm\ is the mean value of the
limit for the ensemble of simulated experiments.

\item
The confidence
level for consistency with the \sm , calculated as the fraction of the
simulated experiments for which the upper limit on the cross-section times 
branching ratio squared is greater than or equal to the value calculated
using the actual data.  In the absence of signal, a CL of 
50\% is expected on average~\footnote{Values of 100\% correspond
to points where there are no candidate events with non-zero \LR\ in the OPAL data.  In this
case, all toy \mc\ experiments will have a value of \signine\ equal to
or (if there are candidates with non-zero \LR ) greater than the value for 
the actual data.}.

\end{enumerate}

Tables~\ref{tab-CL2},~\ref{tab-CL3},~\ref{tab-CL8},~\ref{tab-CL4} and~\ref{tab-CL5}
show the same information for smuons, staus, charginos with two-body decay, 
charginos with three-body decay and charged Higgs, respectively.

{\footnotesize

\begin{table}[htb]
\begin{minipage}[b]{\textwidth}
\centering
\begin{tabular}{||l||c|c|c|c|c|c||}
\hline\hline
 \dm  & \multicolumn{6}{c||}{\msele\ (GeV)\ } \\
\cline{2-7}
 (GeV) &   45 & 55 & 65 & 75 & 85 & 94 \\
\hline\hline
  \multicolumn{7}{||l||}{signal selection efficiency of the general selection at 189~GeV (\%)} \\
\hline
 2    
 & 13.0$\pm$1.1
 & 10.5$\pm$1.0
 &  7.8$\pm$0.8
 &  4.2$\pm$0.6
 &  1.4$\pm$0.4
 &  0.4$\pm$0.2
 \\
 2.5  
 & 25.1$\pm$1.4
 & 21.4$\pm$1.3
 & 22.1$\pm$1.3
 & 18.5$\pm$1.2
 & 12.9$\pm$1.1
 &  7.4$\pm$0.8
 \\
 5    
 & 56.6$\pm$1.6
 & 59.5$\pm$1.6
 & 59.0$\pm$1.6
 & 60.5$\pm$1.5
 & 60.2$\pm$1.5
 & 55.9$\pm$1.6
 \\
 10   
 & 71.0$\pm$1.4
 & 74.4$\pm$1.4
 & 76.7$\pm$1.3
 & 76.3$\pm$1.3
 & 77.5$\pm$1.3
 & 76.0$\pm$1.4
 \\
 20   
 & 78.8$\pm$1.3
 & 81.6$\pm$1.2
 & 85.2$\pm$1.1
 & 85.2$\pm$1.1
 & 84.6$\pm$1.1
 & 85.9$\pm$1.1
 \\
 $m$/2  
 & 78.5$\pm$1.3
 & 84.8$\pm$1.1
 & 87.4$\pm$1.0
 & 90.6$\pm$0.9
 & 90.4$\pm$0.9
 & 92.1$\pm$0.9
 \\
 $m-20$ 
 & 79.0$\pm$1.3
 & 85.6$\pm$1.1
 & 88.8$\pm$1.0
 & 90.4$\pm$0.9
 & 91.8$\pm$0.9
 & 93.1$\pm$0.8
 \\
 $m-10$ 
 & 78.9$\pm$1.3
 & 84.9$\pm$1.1
 & 89.3$\pm$1.0
 & 90.3$\pm$0.9
 & 91.1$\pm$0.9
 & 92.5$\pm$0.8
 \\
 $m$    
 & 77.4$\pm$1.3
 & 84.1$\pm$1.2
 & 88.7$\pm$1.0
 & 90.1$\pm$0.9
 & 91.3$\pm$0.9
 & 93.2$\pm$0.8
 \\

\hline\hline
  \multicolumn{7}{||l||}{95\% CL upper limit on the cross-section times $BR^2(\sele \rightarrow \mathrm{e} \nt_1)$ (fb)} \\
\hline
2     
&     93.8
&    117.2
&    164.8
&    305.5
&    973.8
&   4138.6
\\
2.5   
&     49.2
&     56.5
&     56.1
&     67.7
&    106.1
&    223.7
\\
5     
&     36.7
&     25.7
&     24.3
&     21.9
&     23.3
&     29.6
\\
10    
&     63.1
&     50.2
&     43.8
&     30.8
&     26.8
&     22.0
\\
20    
&     64.8
&     50.9
&     44.7
&     48.9
&     54.7
&     25.6
\\
$m$/2 
&     92.3
&     73.2
&     79.4
&     54.5
&     35.3
&     20.0
\\
$m-20$
&     97.8
&    101.5
&     93.3
&     87.8
&     63.4
&     23.5
\\
$m-10$
&     91.5
&     76.9
&     73.8
&     89.9
&     70.4
&     29.6
\\
$m$   
&     81.4
&     84.6
&     85.8
&     86.8
&     64.0
&     34.5
\\

\hline\hline
  \multicolumn{7}{||l||}{expected upper limit on the cross-section times $BR^2(\sele \rightarrow \mathrm{e} \nt_1)$ (fb)} \\
\hline
2     
&    148.3
&    175.1
&    237.0
&    414.5
&   1325.7
&   4616.9
\\
2.5   
&     82.0
&     90.9
&     87.4
&    104.9
&    136.6
&    264.2
\\
5     
&     42.2
&     39.3
&     38.9
&     34.6
&     33.8
&     34.1
\\
10    
&     40.4
&     33.3
&     28.5
&     25.8
&     24.5
&     24.8
\\
20    
&     69.1
&     52.8
&     43.2
&     34.8
&     27.9
&     23.6
\\
$m$/2 
&     76.1
&     71.5
&     62.5
&     62.0
&     57.2
&     33.6
\\
$m-20$
&     80.7
&     78.7
&     79.3
&     80.0
&     70.3
&     38.9
\\
$m-10$
&     84.9
&     78.6
&     79.6
&     81.1
&     69.3
&     37.9
\\
$m$   
&     81.9
&     76.0
&     76.0
&     80.8
&     71.1
&     38.8
\\

\hline\hline
  \multicolumn{7}{||l||}{CL for consistency with SM (\%)} \\
\hline
2     
& 100.0
& 100.0
& 100.0
& 100.0
& 100.0
& 100.0
\\
2.5   
& 100.0
& 100.0
& 100.0
& 100.0
& 100.0
& 100.0
\\
5     
&  57.0
&  88.6
&  90.5
&  95.4
&  89.1
& 100.0
\\
10    
&   5.2
&   6.8
&   6.3
&  18.0
&  33.1
&  83.9
\\
20    
&  47.6
&  44.6
&  38.2
&  10.4
&   0.6
&  19.7
\\
$m$/2 
&  20.7
&  37.7
&  17.9
&  53.6
&  88.8
&  99.9
\\
$m-20$
&  20.3
&  14.4
&  22.8
&  27.8
&  52.0
&  95.0
\\
$m-10$
&  30.5
&  43.2
&  45.6
&  27.4
&  37.2
&  72.3
\\
$m$   
&  38.2
&  28.1
&  24.9
&  30.9
&  51.2
&  56.0
\\

\hline\hline
\end{tabular}
\caption[]{
Signal selection efficiency of the general selection at 189~GeV,
95\% CL upper limit on the cross-section times 
$BR^2(\sele \rightarrow \mathrm{e} \nt_1)$, 
expected upper limit on the cross-section times 
$BR^2(\sele \rightarrow \mathrm{e} \nt_1)$, 
and the confidence level for consistency with the \sm\ in the search for
\selepair\ production for different values of \msele\ and \dm .
}
\label{tab-CL1}
\end{minipage}
\end{table}

\begin{table}[htb]
\begin{minipage}[b]{\textwidth}
\centering
\begin{tabular}{||l||c|c|c|c|c|c||}
\hline\hline
 \dm  & \multicolumn{6}{c||}{\msmu\ (GeV)\ } \\
\cline{2-7}
 (GeV) &   45 & 55 & 65 & 75 & 85 & 94 \\
\hline\hline
  \multicolumn{7}{||l||}{signal selection efficiency of the general selection at 189~GeV (\%)} \\
\hline
 2    
 & 14.9$\pm$1.1
 & 14.1$\pm$1.1
 &  9.8$\pm$0.9
 &  6.5$\pm$0.8
 &  0.7$\pm$0.3
 &  0.0$\pm$0.0
 \\
 2.5  
 & 26.7$\pm$1.4
 & 27.7$\pm$1.4
 & 25.0$\pm$1.4
 & 21.3$\pm$1.3
 & 14.3$\pm$1.1
 &  8.0$\pm$0.9
 \\
 5    
 & 58.7$\pm$1.6
 & 60.5$\pm$1.5
 & 60.4$\pm$1.5
 & 60.0$\pm$1.5
 & 60.3$\pm$1.5
 & 57.4$\pm$1.6
 \\
 10   
 & 75.8$\pm$1.4
 & 76.7$\pm$1.3
 & 76.2$\pm$1.3
 & 76.4$\pm$1.3
 & 76.5$\pm$1.3
 & 74.1$\pm$1.4
 \\
 20   
 & 85.4$\pm$1.1
 & 86.0$\pm$1.1
 & 84.8$\pm$1.1
 & 84.2$\pm$1.2
 & 84.0$\pm$1.2
 & 83.2$\pm$1.2
 \\
 $m$/2  
 & 86.3$\pm$1.1
 & 87.8$\pm$1.0
 & 87.2$\pm$1.1
 & 90.4$\pm$0.9
 & 90.4$\pm$0.9
 & 91.7$\pm$0.9
 \\
 $m-20$ 
 & 86.7$\pm$1.1
 & 89.2$\pm$1.0
 & 89.7$\pm$1.0
 & 91.2$\pm$0.9
 & 92.9$\pm$0.8
 & 92.8$\pm$0.8
 \\
 $m-10$
 & 88.5$\pm$1.0
 & 90.6$\pm$0.9
 & 89.6$\pm$1.0
 & 91.3$\pm$0.9
 & 92.4$\pm$0.8
 & 93.0$\pm$0.8
 \\
 $m$    
 & 88.6$\pm$1.0
 & 90.6$\pm$0.9
 & 89.8$\pm$1.0
 & 91.1$\pm$0.9
 & 92.1$\pm$0.9
 & 92.5$\pm$0.8
 \\

\hline\hline
  \multicolumn{7}{||l||}{95\% CL upper limit on the cross-section times $BR^2(\smu \rightarrow \mu \nt_1)$ (fb)} \\
\hline
2     
&    105.6
&    119.3
&    177.3
&    300.8
&   2069.3
& --
\\
2.5   
&     63.3
&     61.5
&     72.0
&     88.8
&     98.8
&    206.9
\\
5     
&     30.3
&     31.0
&     39.5
&     31.8
&     34.1
&     28.8
\\
10    
&     35.0
&     22.7
&     18.5
&     17.3
&     18.2
&     22.3
\\
20    
&     53.5
&     38.7
&     39.9
&     41.3
&     37.6
&     25.5
\\
$m$/2 
&     64.7
&     76.7
&     81.6
&     71.1
&     51.7
&     44.1
\\
$m-20$
&     80.1
&    100.5
&    104.6
&     89.6
&     45.3
&     34.3
\\
$m-10$
&     85.4
&     94.6
&     92.8
&     62.9
&     43.4
&     34.3
\\
$m$   
&     83.5
&     87.7
&     88.8
&     58.9
&     43.5
&     34.7
\\

\hline\hline
  \multicolumn{7}{||l||}{expected upper limit on the cross-section times $BR^2(\smu \rightarrow \mu \nt_1)$ (fb)} \\
\hline
2     
&    112.8
&    116.1
&    169.0
&    286.8
&   2977.5
& --
\\
2.5   
&     67.6
&     64.7
&     68.9
&     82.4
&    119.5
&    226.1
\\
5     
&     34.8
&     33.8
&     31.2
&     28.2
&     28.8
&     30.2
\\
10    
&     30.8
&     27.4
&     25.3
&     22.6
&     22.6
&     24.3
\\
20    
&     47.5
&     39.3
&     35.9
&     31.3
&     25.3
&     22.1
\\
$m$/2 
&     50.9
&     50.9
&     50.4
&     51.0
&     50.1
&     33.4
\\
$m-20$
&     51.8
&     52.8
&     55.8
&     58.7
&     56.8
&     36.4
\\
$m-10$
&     53.1
&     51.1
&     53.7
&     56.5
&     56.5
&     38.1
\\
$m$   
&     50.6
&     49.7
&     52.9
&     57.6
&     56.7
&     39.5
\\

\hline\hline
  \multicolumn{7}{||l||}{CL for consistency with SM (\%)} \\
\hline
2     
&  64.6
&  47.9
&  45.7
&  43.7
& 100.0
& --
\\
2.5   
&  65.7
&  60.6
&  54.1
&  40.2
& 100.0
& 100.0
\\
5     
&  57.8
&  47.5
&  19.0
&  35.9
&  27.2
& 100.0
\\
10    
&  26.1
&  63.8
&  76.1
& 100.0
& 100.0
& 100.0
\\
20    
&  27.8
&  42.1
&  30.0
&  15.0
&   7.5
&   8.5
\\
$m$/2 
&  17.6
&   7.6
&   4.8
&  10.6
&  36.0
&  14.2
\\
$m-20$
&   6.3
&   1.7
&   1.1
&   6.1
&  66.8
&  49.0
\\
$m-10$
&   4.4
&   1.3
&   2.7
&  29.1
&  69.0
&  52.3
\\
$m$   
&   3.4
&   2.8
&   3.9
&  36.3
&  71.3
&  57.1
\\

\hline\hline
\end{tabular}
\caption[]{
Signal selection efficiency of the general selection at 189~GeV,
95\% CL upper limit on the cross-section times 
$BR^2(\smu \rightarrow \mu \nt_1)$, 
expected upper limit on the cross-section times 
$BR^2(\smu \rightarrow \mu \nt_1)$, 
and the confidence level for consistency with the \sm\ in the search for
\smupair\ production for different values of \msmu\ and \dm .
}
\label{tab-CL2}
\end{minipage}
\end{table}

\begin{table}[htb]
\begin{minipage}[b]{\textwidth}
\centering
\begin{tabular}{||l||c|c|c|c|c|c||}
\hline\hline
 \dm  & \multicolumn{6}{c||}{\mstau\ (GeV)\ } \\
\cline{2-7}
 (GeV) &   45 & 55 & 65 & 75 & 85 & 94 \\
\hline\hline
  \multicolumn{7}{||l||}{signal selection efficiency of the general selection at 189~GeV (\%)} \\
\hline
 2    
 &  0.2$\pm$0.0
 &  0.1$\pm$0.0
 &  0.1$\pm$0.0
 &  0.0$\pm$0.0
 &  0.0$\pm$0.0
 &  0.0$\pm$0.0
 \\
 2.5  
 &  0.7$\pm$0.1
 &  0.5$\pm$0.1
 &  0.3$\pm$0.0
 &  0.3$\pm$0.0
 &  0.1$\pm$0.0
 &  0.0$\pm$0.0
 \\
 5    
 & 15.6$\pm$0.3
 & 14.0$\pm$0.3
 & 13.1$\pm$0.3
 & 11.6$\pm$0.3
 & 10.3$\pm$0.3
 &  9.1$\pm$0.3
 \\
 10   
 & 38.3$\pm$0.4
 & 39.1$\pm$0.4
 & 39.5$\pm$0.4
 & 38.8$\pm$0.4
 & 39.4$\pm$0.4
 & 38.8$\pm$0.4
 \\
 20   
 & 57.4$\pm$0.4
 & 59.4$\pm$0.4
 & 59.9$\pm$0.4
 & 60.9$\pm$0.4
 & 60.9$\pm$0.4
 & 62.2$\pm$0.4
 \\
 $m$/2  
 & 59.3$\pm$0.4
 & 64.8$\pm$0.4
 & 69.6$\pm$0.4
 & 71.8$\pm$0.4
 & 73.7$\pm$0.4
 & 74.5$\pm$0.4
 \\
 $m-20$ 
 & 60.6$\pm$0.4
 & 69.1$\pm$0.4
 & 73.4$\pm$0.4
 & 74.4$\pm$0.4
 & 77.5$\pm$0.4
 & 78.3$\pm$0.4
 \\
 $m-10$ 
 & 65.2$\pm$0.4
 & 71.1$\pm$0.4
 & 73.9$\pm$0.4
 & 76.2$\pm$0.4
 & 77.7$\pm$0.4
 & 79.2$\pm$0.4
 \\
 $m$    
 & 66.2$\pm$0.4
 & 71.3$\pm$0.4
 & 74.5$\pm$0.4
 & 76.1$\pm$0.4
 & 77.7$\pm$0.4
 & 79.0$\pm$0.4
 \\

\hline\hline
  \multicolumn{7}{||l||}{95\% CL upper limit on the cross-section times $BR^2(\stau \rightarrow \tau \nt_1)$ (fb)} \\
\hline
2     
&  12180.7
&  32560.6
&  16554.6
& --
& --
& --
\\
2.5   
&   2678.7
&   4145.1
&   5526.8
&   6808.6
&  19454.0
& --
\\
5     
&    137.6
&    168.4
&    181.7
&    243.9
&    274.5
&    324.1
\\
10    
&    106.8
&     92.1
&     81.5
&     75.7
&     74.2
&     86.7
\\
20    
&    101.1
&     92.8
&     84.7
&     86.5
&     77.1
&     59.2
\\
$m$/2 
&    100.8
&     97.5
&     87.4
&     76.3
&     76.2
&     90.6
\\
$m-20$
&    106.4
&     93.5
&     78.7
&     75.8
&     76.0
&     87.6
\\
$m-10$
&    119.8
&     97.1
&     83.2
&     75.5
&     73.7
&     97.6
\\
$m$   
&    115.4
&     94.1
&     86.1
&     76.6
&     81.3
&     87.6
\\

\hline\hline
  \multicolumn{7}{||l||}{expected upper limit on the cross-section times $BR^2(\stau \rightarrow \tau \nt_1)$ (fb)} \\
\hline
2     
&  11163.9
&  20013.4
&  19897.3
& --
& --
& --
\\
2.5   
&   2916.5
&   4380.8
&   6850.2
&   7709.0
&  17774.8
& --
\\
5     
&    189.5
&    209.3
&    221.8
&    237.2
&    262.6
&    284.6
\\
10    
&    107.0
&    103.4
&     97.4
&     98.2
&     94.2
&     92.1
\\
20    
&    103.4
&     95.8
&     89.7
&     82.9
&     79.1
&     74.3
\\
$m$/2 
&    106.9
&    104.7
&     99.5
&     95.1
&     93.8
&     96.1
\\
$m-20$
&    108.3
&    107.1
&    108.3
&    107.5
&    108.1
&    112.2
\\
$m-10$
&    113.9
&    112.1
&    114.4
&    109.7
&    114.4
&    111.6
\\
$m$   
&    118.2
&    114.1
&    112.4
&    110.6
&    113.7
&    114.0
\\

\hline\hline
  \multicolumn{7}{||l||}{CL for consistency with SM (\%)} \\
\hline
2     
&  37.4
&   7.4
& 100.0
& --
& --
& --
\\
2.5   
&  56.8
&  41.2
&  77.2
&  65.5
&  45.3
& --
\\
5     
&  75.5
&  66.6
&  62.8
&  38.6
&  32.8
&  23.0
\\
10    
&  40.7
&  56.8
&  61.3
&  70.0
&  66.6
&  50.3
\\
20    
&  43.5
&  44.0
&  46.9
&  37.7
&  42.5
&  67.2
\\
$m$/2 
&  46.1
&  47.5
&  54.7
&  64.4
&  64.1
&  47.5
\\
$m-20$
&  41.4
&  54.3
&  74.1
&  76.0
&  75.0
&  67.0
\\
$m-10$
&  33.0
&  56.0
&  72.9
&  79.0
&  83.2
&  54.0
\\
$m$   
&  40.6
&  59.7
&  67.4
&  79.0
&  73.3
&  69.6
\\

\hline\hline
\end{tabular}
\caption[]{
Signal selection efficiency of the general selection at 189~GeV,
95\% CL upper limit on the cross-section times 
$BR^2(\stau \rightarrow \tau \nt_1)$, 
expected upper limit on the cross-section times 
$BR^2(\stau \rightarrow \tau \nt_1)$, 
and the confidence level for consistency with the \sm\ in the search for
\staupair\ production for different values of \mstau\ and \dm .
}
\label{tab-CL3}
\end{minipage}
\end{table}

\begin{table}[htb]
\begin{minipage}[b]{\textwidth}
\centering
\begin{tabular}{||l||c|c|c|c|c|c||}
\hline\hline
 \dm  & \multicolumn{6}{c||}{\mch\ (GeV)\ } \\
\cline{2-7}
 (GeV) &   50 & 60 & 70 & 80 & 90 & 94 \\
\hline\hline
  \multicolumn{7}{||l||}{signal selection efficiency of the general selection at 189~GeV (\%)} \\
\hline
 2    
 &  7.9$\pm$0.4
 &  6.6$\pm$0.4
 &  4.9$\pm$0.3
 &  2.3$\pm$0.2
 &  0.5$\pm$0.1
 &  0.1$\pm$0.0
 \\
 3    
 & 21.6$\pm$0.7
 & 21.6$\pm$0.7
 & 20.6$\pm$0.6
 & 19.5$\pm$0.6
 & 15.8$\pm$0.6
 & 13.3$\pm$0.5
 \\
 4    
 & 34.3$\pm$0.8
 & 33.5$\pm$0.7
 & 33.7$\pm$0.7
 & 33.6$\pm$0.7
 & 31.5$\pm$0.7
 & 32.5$\pm$0.7
 \\
 5    
 & 41.3$\pm$0.8
 & 43.4$\pm$0.8
 & 42.5$\pm$0.8
 & 41.9$\pm$0.8
 & 43.0$\pm$0.8
 & 44.7$\pm$0.8
 \\
 10   
 & 61.0$\pm$0.8
 & 64.4$\pm$0.8
 & 63.6$\pm$0.8
 & 63.9$\pm$0.8
 & 63.5$\pm$0.8
 & 65.4$\pm$0.8
 \\
 20   
 &  --
 & 75.8$\pm$0.7
 & 77.4$\pm$0.7
 & 78.2$\pm$0.7
 & 79.5$\pm$0.6
 & 78.3$\pm$0.7
 \\
 \mf\ 
 &  --
 & 77.1$\pm$0.7
 & 82.2$\pm$0.6
 & 82.7$\pm$0.6
 & 85.8$\pm$0.6
 & 86.6$\pm$0.5
 \\
 $m-35$ 
 & 69.7$\pm$0.7
 & 78.3$\pm$0.7
 & 83.2$\pm$0.6
 & 85.9$\pm$0.6
 & 87.4$\pm$0.5
 & 89.0$\pm$0.5
 \\

\hline\hline
  \multicolumn{7}{||l||}{95\% CL upper limit on the cross-section times 
$BR^2(\chpm \rightarrow \ell^\pm \snu)$ (fb)} \\
\hline
2     
&    241.3
&    262.3
&    330.0
&    755.6
&   3368.0
&  20947.5
\\
3     
&    103.8
&     89.2
&     98.6
&    103.8
&     88.6
&    125.3
\\
4     
&     62.3
&     60.7
&     68.3
&     61.5
&     80.1
&     54.7
\\
5     
&     53.0
&     44.9
&     58.5
&     50.3
&     63.6
&     62.5
\\
10    
&     92.2
&     82.4
&     57.7
&     39.0
&     39.4
&     38.2
\\
20    
& --
&     72.1
&     64.8
&     60.8
&     65.3
&     40.8
\\
\mf\  
& --
&     81.9
&     83.9
&     88.2
&     61.5
&     53.7
\\
$m-35$
&     84.6
&     87.0
&    116.9
&    118.5
&     70.1
&     42.6
\\

\hline\hline
  \multicolumn{7}{||l||}{expected upper limit on the cross-section times 
$BR^2(\chpm \rightarrow \ell^\pm \snu)$ (fb)} \\
\hline
2     
&    289.4
&    313.9
&    387.6
&    900.9
&   4055.8
&  22119.2
\\
3     
&    123.1
&    117.9
&    108.4
&    119.4
&    134.2
&    165.3
\\
4     
&     86.7
&     82.2
&     75.4
&     70.9
&     77.4
&     73.1
\\
5     
&     77.5
&     68.4
&     68.7
&     64.7
&     59.0
&     60.1
\\
10    
&     79.5
&     64.0
&     58.2
&     52.1
&     47.8
&     44.3
\\
20    
& --
&    108.5
&     84.7
&     68.2
&     49.7
&     38.7
\\
\mf\  
& --
&    121.0
&    119.0
&    115.5
&    108.3
&     74.7
\\
$m-35$
&    103.8
&    130.8
&    154.0
&    166.4
&    143.7
&     88.8
\\

\hline\hline
  \multicolumn{7}{||l||}{CL for consistency with SM (\%)} \\
\hline
2     
&  63.3
&  59.1
&  68.3
&  70.7
&  76.5
&  66.1
\\
3     
&  61.4
&  72.3
&  48.1
&  56.9
& 100.0
&  73.5
\\
4     
&  80.1
&  77.6
&  51.8
&  53.4
&  36.9
&  70.2
\\
5     
&  82.9
&  88.6
&  59.0
&  72.0
&  33.5
&  29.5
\\
10    
&  24.8
&  14.7
&  42.1
&  76.4
&  64.2
&  59.2
\\
20    
& --
&  80.9
&  70.0
&  53.1
&  15.2
&  33.6
\\
\mf\  
& --
&  79.5
&  76.0
&  69.4
&  93.4
&  79.0
\\
$m-35$
&  63.9
&  81.4
&  63.6
&  72.7
&  97.0
&  98.9
\\

\hline\hline
\end{tabular}
\caption[]{
Signal selection efficiency of the general selection at 189~GeV,
95\% CL upper limit on the cross-section times 
$BR^2(\chpm \rightarrow \ell^\pm \snu)$, 
expected upper limit on the cross-section times 
$BR^2(\chpm \rightarrow \ell^\pm \snu)$, 
and the confidence level for consistency with the \sm\ in the search for
\chargtwo\ production for different values of \mch\ and \dm .  The
bins without entries correspond to values of $m_{\snu}<35$, which are
excluded and therefore not considered in the analysis. 
}
\label{tab-CL8}
\end{minipage}
\end{table}

\begin{table}[htb]
\begin{minipage}[b]{\textwidth}
\centering
\begin{tabular}{||l||c|c|c|c|c|c||}
\hline\hline
 \dm  & \multicolumn{6}{c||}{\mch\ (GeV)\ } \\
\cline{2-7}
 (GeV) &   50 & 60 & 70 & 80 & 90 & 94 \\
\hline\hline
  \multicolumn{7}{||l||}{signal selection efficiency of the general selection at 189~GeV (\%)} \\
\hline
 3    
 &  3.7$\pm$0.3
 &  3.2$\pm$0.3
 &  2.5$\pm$0.2
 &  1.0$\pm$0.2
 &  0.2$\pm$0.1
 &  0.2$\pm$0.1
 \\
 5    
 & 17.1$\pm$0.6
 & 16.3$\pm$0.6
 & 15.2$\pm$0.6
 & 13.3$\pm$0.5
 & 11.4$\pm$0.5
 & 10.8$\pm$0.5
 \\
 10   
 & 39.9$\pm$0.8
 & 40.8$\pm$0.8
 & 40.9$\pm$0.8
 & 41.6$\pm$0.8
 & 40.8$\pm$0.8
 & 41.6$\pm$0.8
 \\
 20   
 & 59.1$\pm$0.8
 & 60.4$\pm$0.8
 & 62.2$\pm$0.8
 & 60.8$\pm$0.8
 & 62.8$\pm$0.8
 & 63.8$\pm$0.8
 \\
 $m$/2  
 & 63.3$\pm$0.8
 & 68.7$\pm$0.7
 & 74.3$\pm$0.7
 & 77.9$\pm$0.7
 & 79.3$\pm$0.6
 & 80.3$\pm$0.6
 \\
 $m-20$ 
 & 67.3$\pm$0.7
 & 72.6$\pm$0.7
 & 78.0$\pm$0.7
 & 81.2$\pm$0.6
 & 83.9$\pm$0.6
 & 84.0$\pm$0.6
 \\
 $m-10$ 
 & 70.1$\pm$0.7
 & 75.9$\pm$0.7
 & 79.4$\pm$0.6
 & 83.0$\pm$0.6
 & 87.0$\pm$0.5
 & 88.1$\pm$0.5
 \\
 $m$    
 & 73.2$\pm$0.7
 & 77.4$\pm$0.7
 & 81.2$\pm$0.6
 & 85.2$\pm$0.6
 & 88.1$\pm$0.5
 & 89.5$\pm$0.5
 \\

\hline\hline
  \multicolumn{7}{||l||}{95\% CL upper limit on the cross-section times 
$BR^2(\chpm \rightarrow \ell^\pm \nu \chz)$ (fb)} \\
\hline
3     
&    355.9
&    540.9
&    598.9
&   1578.0
&   5613.2
&  11457.9
\\
5     
&    111.0
&    116.1
&    134.6
&    114.5
&    156.9
&    272.9
\\
10    
&     74.7
&     60.4
&     66.5
&     61.0
&     56.2
&     65.5
\\
20    
&     91.3
&     78.4
&     68.0
&     64.3
&     51.0
&     52.8
\\
$m$/2 
&    105.8
&     99.7
&     82.9
&     69.9
&     88.0
&    121.6
\\
$m-20$
&    118.5
&     85.2
&     91.4
&     91.8
&     89.2
&     89.4
\\
$m-10$
&    111.6
&    103.7
&    104.2
&    101.6
&     88.1
&     60.6
\\
$m$   
&    133.7
&    132.4
&    117.3
&    117.2
&     99.6
&     57.6
\\

\hline\hline
  \multicolumn{7}{||l||}{expected upper limit on the cross-section times
$BR^2(\chpm \rightarrow \ell^\pm \nu \chz)$ (fb)} \\
\hline
3     
&    563.5
&    668.8
&    791.2
&   2050.1
&   8303.3
&  10493.8
\\
5     
&    163.4
&    154.9
&    165.4
&    169.6
&    203.6
&    222.4
\\
10    
&     90.4
&     82.6
&     80.5
&     76.1
&     76.4
&     74.4
\\
20    
&     92.0
&     81.2
&     70.0
&     67.3
&     63.0
&     61.7
\\
$m$/2 
&    100.3
&     94.4
&     89.9
&     81.2
&     78.5
&     82.5
\\
$m-20$
&    103.7
&    114.8
&    114.9
&    113.9
&    131.0
&    140.5
\\
$m-10$
&    128.2
&    130.9
&    135.3
&    136.9
&    166.9
&    136.5
\\
$m$   
&    139.1
&    144.8
&    154.9
&    183.3
&    179.4
&    133.9
\\

\hline\hline
  \multicolumn{7}{||l||}{CL for consistency with SM (\%)} \\
\hline
3     
&  94.9
&  62.7
&  83.3
&  76.7
& 100.0
&  37.5
\\
5     
&  82.9
&  72.6
&  62.5
&  88.5
&  79.9
&  30.5
\\
10    
&  62.1
&  76.4
&  62.6
&  68.0
&  74.8
&  53.0
\\
20    
&  41.3
&  44.3
&  43.7
&  46.3
&  65.2
&  60.2
\\
$m$/2 
&  35.7
&  35.7
&  49.8
&  54.9
&  28.6
&   8.4
\\
$m-20$
&  26.7
&  69.0
&  65.0
&  62.3
&  80.5
&  84.9
\\
$m-10$
&  52.9
&  63.4
&  66.8
&  69.9
&  94.2
&  99.0
\\
$m$   
&  40.9
&  43.9
&  64.9
&  78.6
&  89.8
&  99.0
\\

\hline\hline
\end{tabular}
\caption[]{
Signal selection efficiency of the general selection at 189~GeV,
95\% CL upper limit on the cross-section times 
$BR^2(\chpm \rightarrow \ell^\pm \nu \chz)$, 
expected upper limit on the cross-section times 
$BR^2(\chpm \rightarrow \ell^\pm \nu \chz)$, 
and the confidence level for consistency with the \sm\ in the search for
\chargthree\ production for different values of \mch\ and \dm .
}
\label{tab-CL4}
\end{minipage}
\end{table}

\begin{table}[htb]
\begin{minipage}[b]{\textwidth}
\centering
\begin{tabular}{||c|c|c|c|c|c||}
\hline\hline
  \multicolumn{6}{||c||}{\mH\ (GeV)\ } \\
\cline{1-6}
  45 & 55 & 65 & 75 & 85 & 94 \\
\hline\hline
  \multicolumn{6}{||l||}{signal selection efficiency of the general selection at 189~GeV (\%)} \\
\hline
      
   66.6$\pm$0.7
 & 72.3$\pm$0.7
 & 74.6$\pm$0.7
 & 76.9$\pm$0.7
 & 77.6$\pm$0.7
 & 79.7$\pm$0.6
 \\

\hline\hline
  \multicolumn{6}{||l||}{95\% CL upper limit on the cross-section times 
$BR^2(\dH)$ (fb)} \\
\hline
     113.6
&     84.8
&     81.5
&     73.2
&     77.0
&     80.1
\\

\hline\hline
  \multicolumn{6}{||l||}{expected upper limit on the cross-section times 
$BR^2(\dH)$ (fb)} \\
\hline
     117.7
&    113.2
&    114.0
&    114.0
&    112.3
&    111.4
\\

\hline\hline
  \multicolumn{6}{||l||}{CL for consistency with SM (\%)} \\
\hline
   40.8
&  70.3
&  75.0
&  83.5
&  78.7
&  76.8
\\

\hline\hline
\end{tabular}
\caption[]{
Signal selection efficiency of the general selection at 189~GeV,
95\% CL upper limit on the cross-section times $BR^2(\dH)$, 
expected upper limit on the cross-section times $BR^2(\dH)$, 
and the confidence level for consistency with the \sm\ in the search for
$\mathrm{H}^+\mathrm{H}^-$ production for different values of \mH .
}
\label{tab-CL5}
\end{minipage}
\end{table}

}

For some points in $m$ and \dm\ in Tables~\ref{tab-CL1} and~\ref{tab-CL2}, the
confidence level for consistency with the \sm\ is small (around 1\%).  
The probability of getting a low confidence level for one or more points in 
$m$ and \dm\ for one or more of the search channels depends on
the degree of correlation among the different ($m$, \dm) points
and among the different channels.  The degree of correlations between
adjacent points is strong when the momentum distributions for those points are
similar.  The momentum distributions vary slowly with
both $m$ and \dm\ when \dm\ is high (hence the clustering of low
confidence level values in Table~\ref{tab-CL2}), but vary considerably
with \dm\ when \dm\ is low.

This effect was investigated
by calculating the cross-section limits for each of 1000 experiments in which 
the data is simulated by randomly selected \smc\ events.  For each experiment, 
the number of events taken from a
\mc\ sample simulating a given process is drawn from a Poisson
distribution with mean equal to the number of events expected for that process.
For each experiment, the confidence level at each ($m$, \dm) point at
which signal \mc\ has been generated was 
calculated as already described, and the number
of experiments for which a confidence level of 0.6\%
\footnote{This is the lowest value of the confidence level in Tables~\ref{tab-CL1} 
to~\ref{tab-CL5}.} or less is obtained for 
at least one point in $m$ and \dm\ in at least one search channel was 
determined.  This was found to be the case for 390 of the 1000 experiments.

As a cross-check, taking the mean of the ensemble of limits obtained at each 
($m$, \dm) point for these simulated experiments was used as an alternative to 
the method described above to obtain \expsig .  The results were found to be consistent.

\subsection{Limits on New Particle Masses}
\label{sec:mass}

\begin{figure}[htbp]
 \epsfxsize=\textwidth 
 \epsffile[0 0 580 600]{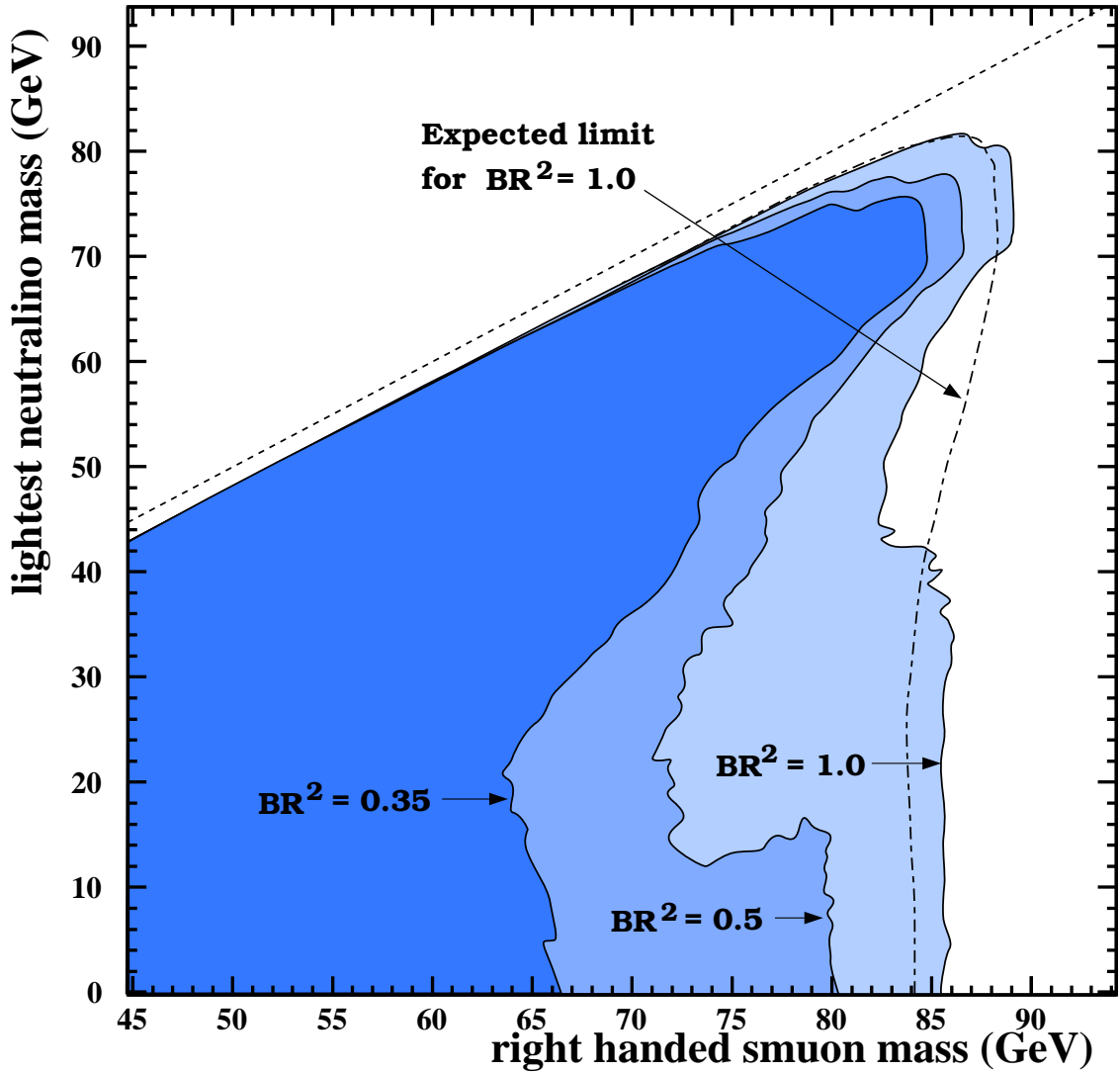}
 \caption{\sl 
95\% CL exclusion region for right-handed smuon pair 
production obtained by combining the $\protect\sqrt{s} = 183$ and 189~GeV 
data-sets.
The limits are calculated for several values of
the branching ratio squared for 
$\smu^\pm_R \rightarrow  {\mu^\pm} \nt_1$ that are indicated in the figure.
Otherwise they have no supersymmetry model assumptions.
The kinematically allowed region is indicated by the dashed line.  The
expected limit for BR$^2$~=~1.0, calculated from \mc\ alone, is indicated by
the dash-dotted line.
} 
\label{fig-mssm_2}
\end{figure}

\begin{figure}[htbp]
 \epsfxsize=\textwidth 
 \epsffile[0 0 580 600]{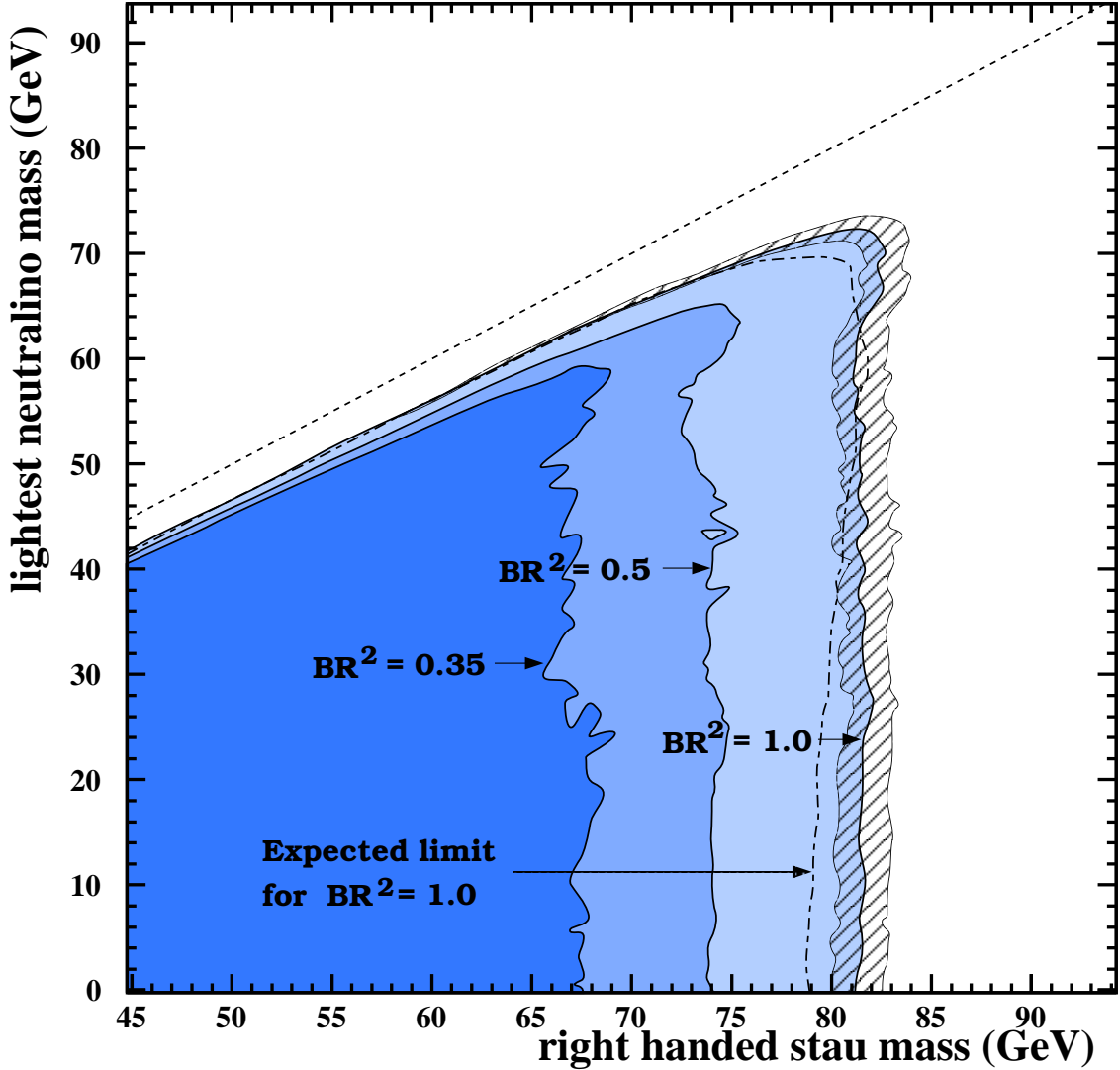}
 \caption{\sl 
95\% CL exclusion region for right-handed stau pair 
production obtained by combining the $\protect\sqrt{s} = 183$ and 189~GeV 
data-sets.
The limits are calculated for several values of
the branching ratio squared for 
$\stau^\pm_R \rightarrow  {\tau^\pm} \nt_1$.
The selection efficiency for \staupair\ is calculated for the case
that the decay \dstau\ produces unpolarised $\tau^\pm$.
Otherwise the limits have no supersymmetry model assumptions.
The hatched area shows the region in which the limit for BR$^2$~=~1.0 can
vary if stau mixing occurs (see text).
The kinematically allowed region is indicated by the dashed
line.  The expected limit for BR$^2$~=~1.0, 
calculated from \mc\ alone, is indicated by
the dash-dotted line.
} 
\label{fig-mssm_3}
\end{figure}

We can use our data to set limits on the masses of right-handed 
sleptons\footnote{
The right-handed slepton is expected to be lighter
than the left-handed slepton. The
right-handed one tends (not generally valid for selectrons)
to
have a lower pair production cross-section, and so
conventionally limits are given for this (usually) conservative case.}
based on the expected right-handed slepton pair cross-sections and
branching ratios.
The cross-sections have been calculated using {\sc Susygen}
at each centre-of-mass energy 
and take into account initial state radiation.
In Figure~\ref{fig-mssm_2} we show  limits on right-handed smuons
 as a function of smuon mass and lightest
neutralino mass for several assumed values of
the branching ratio squared for $\smu^\pm_R \rightarrow  {\mu^\pm} \nt_1$.
The expected limit, calculated using \mc\ only, for a branching ratio
of 100\% is also shown.
For a branching ratio $\smu^\pm_R \rightarrow  {\mu^\pm} \nt_1$ of
100\% and for a smuon-neutralino mass difference exceeding 3~GeV,
right-handed smuons are excluded at 95\% CL for 
masses below 82.3~GeV.
The  95\% CL upper limit on the  production of
right-handed \staupair\ times the
 branching ratio squared for $\stau^\pm_R \rightarrow  {\tau^\pm} \nt_1$
is shown in Figure~\ref{fig-mssm_3}.  The expected limit for a branching ratio
of 100\% is also shown.
For a branching ratio $\stau^\pm_R \rightarrow  {\tau^\pm} \nt_1$ of
100\% and for a stau-neutralino mass difference exceeding 8~GeV,
right-handed staus are excluded at 95\% CL for 
masses below 81.0~GeV.  No mixing between $\stau_L$ and $\stau_R$ is assumed.
However, the cross-section ratio 
$\sigma_{\stau_1^+ \stau_1^-}/\sigma_{\stau_R^+ \stau_R^-}$ at
$\roots$=189~GeV varies 
between 0.89 and 1.20, depending only on the mixing angle.  Using this
information, the limits shown in 
Figure~\ref{fig-mssm_3} can be applied to any degree of stau mixing by 
multiplying the predicted cross-section for $\stau_R^+ \stau_R^-$ by the 
value of $\sigma_{\stau_1^+ \stau_1^-}/\sigma_{\stau_R^+ \stau_R^-}$ 
corresponding to the mixing angle considered.  The hatched region in 
Figure~\ref{fig-mssm_3} shows the range of possible positions of the line 
defining the excluded region for a branching ratio 
$\stau^\pm_1 \rightarrow  {\tau^\pm} \nt_1$ of
100\% for any degree of stau mixing.

For the case of a massless neutralino (or gravitino) and 100\%
branching ratio, right-handed smuons and staus are excluded at 95\% CL for 
masses below 85.4~GeV and 81.1~GeV, respectively, and $\stau^\pm_1$ is
excluded at 95\% CL for masses below 80.0~GeV, for any degree of stau mixing.

\begin{figure}[htbp]
 \epsfxsize=\textwidth 
 \epsffile[0 0 580 600]{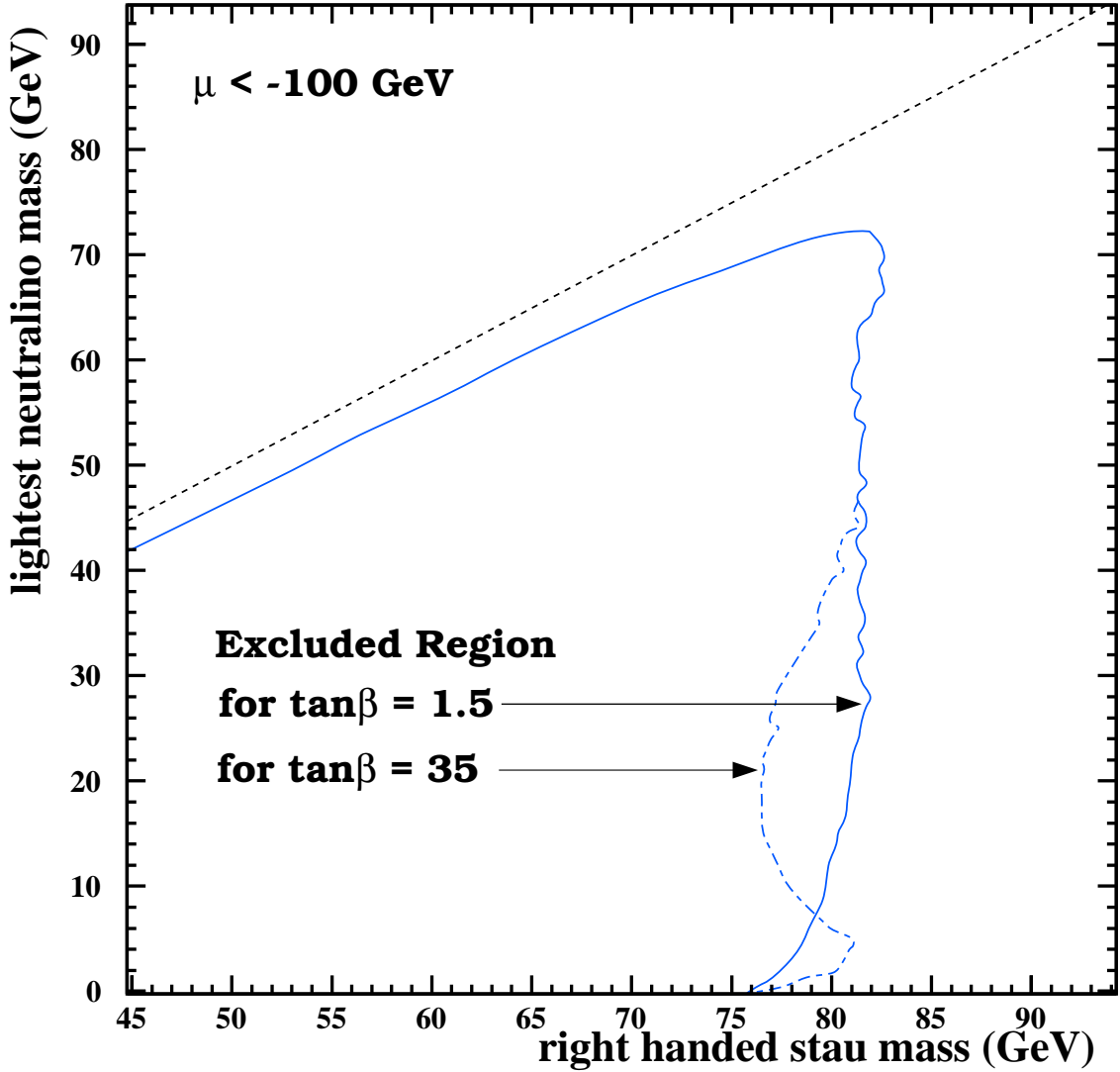}
 \caption{\sl 
For two values of $\tan{\beta}$ and $\mu < -100$~GeV,
 95\% CL exclusion regions for right-handed stau pairs within the MSSM,
obtained by combining the $\protect\sqrt{s} = 183$ and 189~GeV data-sets.
The excluded regions are calculated 
taking into account the 
predicted branching ratio for $\stau^\pm_R \rightarrow  {\tau^\pm}
\nt_1$.
The  gauge unification relation,
$M_1 =  \frac{5}{3} \tan^2 \theta_W M_2$, is assumed in calculating the
MSSM branching ratios.
The selection efficiency for \staupair\ is calculated for the case
that the decay \dstau\ produces unpolarised $\tau^\pm$.
The kinematically allowed region is indicated by the dashed line.
} 
\label{fig-mssm_3a}
\end{figure}

An alternative approach is to set limits
taking into account the 
predicted cross-section and  branching ratio 
for specific choices of the parameters within the 
Minimal Supersymmetric Standard Model (MSSM)\footnote{
In particular 
 regions of the MSSM parameter space, the branching ratio for 
$\sell^\pm \rightarrow  {\ell^\pm} \nt_1$  
can be essentially zero and so 
it is not possible to provide general limits on sleptons within the MSSM
on the basis of this search alone.
The predicted cross-sections and  branching ratios within the MSSM 
are obtained using {\sc Susygen}
and are calculated with the gauge unification relation,
$M_1 =  \frac{5}{3} \tan^2 \theta_W M_2$.}.
For $\mu < -100$~GeV and for two
values of $\tan{\beta}$ (1.5 and 35),
Figures~\ref{fig-mssm_1},~\ref{fig-mssm_2a} and~\ref{fig-mssm_3a}
 show 95\% CL exclusion regions 
in the ($m_{\tilde{\ell}^\pm_{\mathrm{R}}}$, $m_{\nt_1}$) 
plane
for right-handed selectrons, smuons and staus, respectively.
For $\mu < -100$~GeV and $\tan{\beta}=1.5$, right-handed sleptons are 
excluded at
95\% CL as follows:
selectrons with masses below 87.1~GeV for \mbox{$\msele - \mchz > 5$}~GeV;
smuons with masses below 81.7~GeV for \mbox{$\msmu - \mchz > 3$}~GeV;
and staus with masses below 75.9~GeV for \mbox{$\mstau - \mchz > 7$}~GeV.

\section{Summary and Conclusions}
 
A selection of di-lepton events with significant missing transverse momentum 
is performed using a total data sample of 237.4~pb$^{-1}$
at e$^+$e$^-$ centre-of-mass energies of 183 and 189~GeV.
The observed numbers of events, 78 at 183 GeV and 301
at 189~GeV,
are consistent with the numbers expected from Standard Model processes, 
dominantly arising from \wpair\ production with each W decaying
leptonically.

Further discrimination techniques are employed to search for the pair 
production of charged scalar leptons, leptonically decaying charged Higgs 
bosons and charginos that decay to produce a charged lepton and invisible 
particles.
No evidence for new phenomena is apparent and model independent limits 
on the production cross-section times branching ratio squared
for each new physics process are presented.

Assuming a 100\% branching ratio for the decay
$\sell^\pm_R \rightarrow  {\ell^\pm} \nt_1$, we exclude at 95\% CL:
right-handed smuons with masses below 82.3~GeV for 
\mbox{$\msmu - \mchz > 3$}~GeV and
right-handed staus with masses below 81.0~GeV for 
\mbox{$\mstau - \mchz > 8$}~GeV.
Right-handed selectrons are excluded at 95\% CL for 
masses below 87.1~GeV for \mbox{$\msele - \mchz > 5$}~GeV
within the framework of the
 MSSM assuming
$\mu < -100$~GeV and $\tan{\beta} = 1.5$.
Charged Higgs bosons are excluded at 95\% CL for masses below 82.8~GeV,
assuming a 100\% branching ratio for the decay \dH .

The cross-section times branching ratio squared limits 
from the selectron, smuon and two-body chargino searches
presented here 
are used in the interpretation of the results of \cite{OPAL_chargino_189}
in terms of mass limits on charginos and neutralinos.

\bigskip\bigskip
\noindent {\Large\bf Acknowledgements.}

\noindent 
We particularly wish to thank the SL Division for the efficient operation
of the LEP accelerator at all energies
 and for their continuing close cooperation with
our experimental group.  We thank our colleagues from CEA, DAPNIA/SPP,
CE-Saclay for their efforts over the years on the time-of-flight and trigger
systems which we continue to use.  In addition to the support staff at our own
institutions we are pleased to acknowledge the  \\
Department of Energy, USA, \\
National Science Foundation, USA, \\
Particle Physics and Astronomy Research Council, UK, \\
Natural Sciences and Engineering Research Council, Canada, \\
Israel Science Foundation, administered by the Israel
Academy of Science and Humanities, \\
Minerva Gesellschaft, \\
Benoziyo Center for High Energy Physics,\\
Japanese Ministry of Education, Science and Culture (the
Monbusho) and a grant under the Monbusho International
Science Research Program,\\
Japanese Society for the Promotion of Science (JSPS),\\
German Israeli Bi-national Science Foundation (GIF), \\
Bundesministerium f\"ur Bildung, Wissenschaft,
Forschung und Technologie, Germany, \\
National Research Council of Canada, \\
Research Corporation, USA,\\
Hungarian Foundation for Scientific Research, OTKA T-029328, 
T023793 and OTKA F-023259.\\



\newpage

\clearpage

\clearpage

\clearpage

\clearpage

\clearpage

\begin{figure}[htbp]
 \epsfxsize=\textwidth 
 \epsffile[0 0 580 600]{pr290_06.eps_col}
 \caption{\sl
Contours of the 95\% CL upper limits on the selectron pair
cross-section times $BR^2(\sele \rightarrow \mathrm{e} \nt_1)$
at 189 GeV
based on combining the 183 and 189 GeV data-sets 
assuming a $\beta^3/s$ dependence of the cross-section.
The kinematically allowed region is indicated by the dashed line.  The
unshaded region at very low \dm\ is experimentally inaccessible in
this search.
} 
\label{fig:limit_1}
\end{figure}
\clearpage

\begin{figure}[htbp]
 \epsfxsize=\textwidth 
 \epsffile[0 0 580 600]{pr290_07.eps_col}
 \caption{\sl
Contours of the 95\% CL upper limits on the smuon pair
cross-section times $BR^2(\smu \rightarrow \mu \nt_1)$
at 189 GeV
based on combining the 183 and 189 GeV data-sets 
assuming a $\beta^3/s$ dependence of the cross-section.
The kinematically allowed region is indicated by the dashed line.  The
unshaded region at very low \dm\ is experimentally inaccessible in
this search.
} 
\label{fig:limit_2}
\end{figure}
\clearpage

\clearpage

\clearpage

\clearpage

\clearpage

\clearpage

\clearpage

\begin{figure}[htbp]
 \epsfxsize=\textwidth 
 \epsffile[0 0 580 600]{pr290_14.eps_col}
 \caption{\sl 
For two values of $\tan{\beta}$ and $\mu < -100$~GeV,
95\% CL exclusion regions for right-handed selectron pairs within the MSSM, 
obtained by combining the $\protect\sqrt{s} = 183$ and 189~GeV data-sets.
The excluded regions are calculated 
taking into account the 
predicted branching ratio for 
$\sele^\pm_R \rightarrow  {\mathrm{e}^\pm} \nt_1$.
The  gauge unification relation,
$M_1 =  \frac{5}{3} \tan^2 \theta_W M_2$, is assumed in calculating the
MSSM cross-sections and branching ratios.
The kinematically allowed region is indicated by the dashed line.
} 
\label{fig-mssm_1}
\end{figure}
\clearpage

\begin{figure}[htbp]
 \epsfxsize=\textwidth 
 \epsffile[0 0 580 600]{pr290_15.eps_col}
 \caption{\sl 
For two values of $\tan{\beta}$ and $\mu < -100$~GeV,
 95\% CL exclusion regions for right-handed smuon pairs  within the MSSM,
obtained by combining the $\protect\sqrt{s} = 183$ and 189~GeV data-sets.
The excluded regions are calculated 
taking into account the 
predicted branching ratio for $\smu^\pm_R \rightarrow  {\mu^\pm} \nt_1$.
The  gauge unification relation,
$M_1 =  \frac{5}{3} \tan^2 \theta_W M_2$, is assumed in calculating the
MSSM branching ratios.
The kinematically allowed region is indicated by the dashed line.
} 
\label{fig-mssm_2a}
\end{figure}
\clearpage

\clearpage

\end{document}